\documentstyle[11pt,amsfonts]{article}

\newcommand{\be}{\begin{equation}}
\newcommand{\ee}{\end{equation}}
\newcommand{\bea}{\begin{array}}
\newcommand{\ea}{\end{array}}
\newcommand{\beqa}{\begin{eqnarray}}
\newcommand{\eeqa}{\end{eqnarray}}
\newcommand{\bean}{\begin{eqnarray*}}
\newcommand{\eean}{\end{eqnarray*}}

\def\kbar{{\mathchar'26\mkern-9muk}}
\def\BI{{\rm 1\!l}}

\def\up#1{\leavevmode \raise.16ex\hbox{#1}}

\newcommand{\gapproxeq}{\lower
 .7ex\hbox{$\;\stackrel{\textstyle >}{\sim}\;$}}
\newcommand{\lapproxeq}{\lower .7ex\hbox{$\;\stackrel
{\textstyle <}{\sim}\;$}}

\newcounter{appendice}

\def\thebibliography#1{{\bf REFERENCES\markboth
 {REFERENCES}{REFERENCES}}\list
 {[\arabic{enumi}]}{\settowidth\labelwidth{[#1]}\leftmargin\labelwidth
 \advance\leftmargin\labelsep
 \usecounter{enumi}}
 \def\newblock{\hskip .11em plus .33em minus -.07em}
 \sloppy
 \sfcode`\.=1000\relax}

\begin{document}

%\vspace*{5mm}

\centerline{ \LARGE  Non-constant Non-commutativity in $2d$ Field Theories}
\vskip .5cm

\centerline{\LARGE and a New Look at Fuzzy Monopoles}

\vskip 2cm

\centerline{ {\sc   A. Stern }  }

\vskip 1cm
\begin{center}

{ Department of Physics, University of Alabama,\\
Tuscaloosa, Alabama 35487, USA}

\end{center}

\vskip 2cm

\vspace*{5mm}

\normalsize
\centerline{\bf ABSTRACT} 

We write  down scalar field theory  and gauge theory on 
two-dimensional noncommutative spaces ${\cal M}$ with nonvanishing curvature and
non-constant non-commutativity.
Usual dynamics results upon
 taking the limit of ${\cal M}$ going to i) a commutative manifold
 ${\cal M}_0$ having nonvanishing curvature and ii)  the
noncommutative plane.  Our procedure does not require introducing
singular algebraic maps or frame fields.   Rather,
we  exploit the K\"ahler structure in the limit i) and identify the symplectic two-form with the volume
two-form.  As an example, we take ${\cal M}$ to be the
stereographically projected fuzzy sphere, and find magnetic monopole solutions
to the noncommutative Maxwell equations.  Although the magnetic charges are conserved, the
classical theory does not require that they be quantized.  The
noncommutative gauge field strength transforms in the usual manner, but the same
 is not, in general, true for the associated potentials.  We develop  a
 perturbation scheme to obtain the expression for gauge transformations about limits i) and ii).  We also obtain the lowest order
 Seiberg-Witten map to write down corrections to the
 commutative field equations and show that solutions to
  Maxwell theory on ${\cal M}_0$ are stable under inclusion of  lowest
 order noncommutative corrections.  The results are applied to the example of
noncommutative AdS${}^2$.

\vspace*{5mm}

\newpage

\section{Introduction}

\setcounter{equation}{0}

Much work has been carried out for field theories on the
noncommutative plane.  This is the case of
constant  non-commutativity.  On the other hand,  not much is known
for field theories on spaces with   non-constant non-commutativity.
Exceptional cases  are when the non-commutativity is
associated with certain
Lie-algebra structures, such as the case with fuzzy spheres and fuzzy
CP${}^n$ models.  (For a review, see \cite{Balachandran:2005ew}.)  Other  cases have been
discussed in
 \cite{Fosco:2004yz},
\cite{Gayral:2005ih}.
  Among the
obstacles  to constructing field theories on general noncommutative spaces  are problems in
defining differentiation, integration and a Dirac operator.\footnote{
  On the other hand, such problems do not arise if one is only
  interested in doing particle mechancis on these spaces, as for
  example is done in \cite{Balachandran:1983pc}.
}  Recently,  scalar field theory
\cite{Pinzul:2005nh} and  gauge theory\cite{Behr:2003qc} have been
formulated on general
curved noncommutative manifolds.  Although the procedures in \cite{Pinzul:2005nh} and \cite{Behr:2003qc} differ,
both make use of  frame fields induced by a non-constant metric.  They  were associated with algebraic maps to the noncommutative plane in
\cite{Pinzul:2005nh}, and
appear 
  in the definition of derivations in \cite{Behr:2003qc}.
 A loss
of generality may result in the noncommutative theory, however, as frame fields are only defined on local coordinate
patches.     In \cite{Pinzul:2005nh} this resulted in the possibility of singular algebraic maps in the noncommutative theory.  Below we
shall develop an alternative approach for two dimensional noncommutative  scalar and gauge
theories which avoids the introduction of  frame fields.  We
also avoid the problem of defining  noncommutative derivatives by writing the theory
algebraically from the start, expressing the underlying commutative
theory in terms of a Poisson bracket algebra.

  In  writing down 
field theories on some   noncommutative
 space  ${\cal M}$,  we shall require that the results be  consistent with deformations of
known theories.   In particular, we  insist that the field theories
 reduce to the standard form in the limit that  ${\cal M}$  reduces i) to a
commutative manifold  with nonvanishing curvature and ii) to the
noncommutative plane.  As stated above, we  restrict to two dimensional field theories.
In that case we can  exploit the K\"ahler structure of the commutative
space and identify the symplectic two-form with the volume
two-form.     The  Lagrangian densities for the commutative  theories can then be
expressed in terms of Poisson brackets.  In passing to the
noncommutative theory we  simply replace the
Poisson bracket by  an  appropriate commutator and the classical
measure by its noncommutative counterpart.  This can be done without
spoiling the symmetries of the underlying commutative space  ${\cal M}_0$, if there are any,
since the Poisson brackets can be constructed to preserve these symmetries.
The resulting free
noncommutative scalar field  and Maxwell equations have a simple
form.   Concerning the latter, 
there are no propagating degrees of freedom, just as with  commutative electrodynamics in two dimension, and  solutions  are characterized by  a constant flux per unit
area and action per unit area.
The noncommutative  field
strength is covariant with respect to gauge transformations.  However, the corresponding transformations of the
potentials are nontrivial and   geometry dependent.

We shall apply the results to find magnetic monopole solutions on the fuzzy sphere.
Electrodynamics on the fuzzy sphere has been of considerable
interest.\cite{Grosse:1992bm},\cite{Klimcik:1997mg},\cite{Grosse:1998da},\cite{Presnajder:1999ky},\cite{Iso:2001mg},\cite{Karabali:2001te},\cite{Balachandran:2003ay},\cite{Steinacker:2003sd}  Ans\"atse for magnetic monopoles  on the fuzzy sphere
have been proposed\cite{Baez:1998he},  although they were  not
required to be deformations of
monopoles on the commutative sphere.
Noncommutative magnetic monopoles were  expressed using the analogue of embedding
coordinates  in
\cite{Karabali:2001te}, and had the correct commutative limit.  Here we obtain 
magnetic monopoles as solutions to the noncommutative Maxwell equations on the analogue of the projective plane.  A
nonsingular  map from the fuzzy sphere
to the noncommutative projective plane was given in
\cite{Alexanian:2000uz}.    (The coordinate singularity appears  only
in  the commutative limit.)  Since it is nonsingular, one can  express
the noncommutative potentials free of Dirac-string singularities.  We find
that the associated magnetic charges are conserved, although not for topological
reasons.  However they need not be
quantized, at least at the classical level.  Alternatively, we can get charge
quantization, upon imposing additional constraints, but these charges
have a singular commutative limit.

 It is common to  realize the
noncommutative algebra  with a star product.     The
Groenewold-Moyal star product is often used, but since it is associated with  constant non-commutativity, it is not very
convenient to realize the algebra on  ${\cal M}$.  More
appropriate is Konsevich's formality map\cite{Kontsevich:1997vb} which
was utilized  in
\cite{Behr:2003qc}.   Alternatively, we shall
rely upon the star product  developed in \cite{Alexanian:2000uz} which
is based on a nonlinear deformation of  coherent states on the complex plane.\cite{mmsz}
An exact  integral expression for this star product was given, which can
be expanded about either limit i) or ii).  An expansion for the measure can also be
given  by simply demanding that the result satisfies the usual
properties of a trace.  By applying these expansions we get corrections
to the  the scalar and Maxwell actions about the two limits.  Although 
approximation schemes for these actions have been given previously \cite{Pinzul:2005nh},\cite{Behr:2003qc}, the one presented here has
the advantage of simplicity.  Concerning
the scalar field action, we get that the lowest order  effects of 
non-commutativity are obtained by just replacing derivatives on ${\cal M}_0$ by
`covariant' ones.   We also compute lowest order  corrections to the
commutative  expression for gauge transformations of the potentials, and show how to
Seiberg-Witten map\cite{Seiberg:1999vs} these corrections away.
Using the Seiberg-Witten map we can also compute corrections
 to the commutative flux through any region, as
 well as to the Maxwell equations and its solutions.  As expected from
 the exact theory, the flux per
 unit area  and action  per
 unit area  are constants  for the solutions, but
their values are shifted  from the commutative results.
Because the shift is small (i.e. of the order of the non-commutativity
parameter) we say  that solutions to the commutative theory are stable under
inclusion of the noncommutative corrections. As an example, we apply
the techniques to the case where ${\cal M}$ is
 the noncommutative analogue of the Lobachevsky plane.  This space is defined 
by a projection from noncommutative AdS${}^2$.  Here we show
how to obtain corrections to
 the solutions to the commutative free scalar field theory.  The solutions to the
commutative Maxwell equations receive no first order corrections.

We review scalar field theory and gauge field
theory in the commutative limit i) in section 2 and the
noncommutative plane limit ii) in section 3.  Field theories on spaces
with  nonvanishing curvature and
non-constant non-commutativity are described  in section 4.  In section 5
we apply the results to
the fuzzy sphere and obtain the analogue of magnetic monopole
solutions.   The 
 first order corrections away from the two
limits are computed in section 6.  In section 7, we apply the results
to the example of noncommutative Lobachevsky plane.  Brief remarks are
made in section 8.

\section{Curved space -  Commutative theory}
\setcounter{equation}{0}

Here we take advantage of the K\"ahler structure of any two
dimensional commutative manifold ${\cal M}_0$ to express scalar field and
gauge field  Lagrangians on  any
coordinate patch $P_0$ of ${\cal M}_0$   in
terms of Poisson brackets.   Let   $g_{\mu\nu}$ denote the  metric
tensor associated with $P_0$, parameterized by  real coordinates $x^\mu,\;\mu=1,2$.
Alternatively,
we can define complex coordinates $z=x^1+ix^2$ and $\bar
z=x^1-ix^2$.  We introduce a function $\theta_0(z,\bar z)$, which we
assume is nonsingular,  and the
commutative measure  $d\mu_0(z,\bar z)$ on $P_0$  by writing the area  two form    as
\be \sqrt{ g }\; d^2x = \;\frac{i\; dz\wedge d\bar z}{\theta_0(z,\bar z)}\equiv
2\pi\; d\mu_0(z,\bar z)\label{clsclmsr}\; \ee
This can be identified with  a symplectic two-form, with a
corresponding Poisson bracket $\{\;,\;\}$.  So if $\alpha$ and $\beta$ are
functions of $z$ and $\bar z$ their Poisson bracket is 
\be\{\alpha,\beta\} =-i\theta_0(z,\bar z)\;(\partial \alpha\bar\partial
\beta-\bar\partial \alpha\partial
\beta)\;,\label{pbcsct}  \ee where $\partial=\frac\partial{\partial z}$,
$\bar\partial=\frac\partial{\partial \bar z}$.  Its integral over $P_0$ with
respect to the measure  $d\mu_0(z,\bar z)$  vanishes provided $\alpha$ and
$\beta$ vanish sufficiently rapidly as the boundary of $P_0$ is
approached.  More generally, for some region
region $\sigma$ in $P_0$, the integral is equal to boundary terms:
\beqa  \int_\sigma d\mu_0(z,\bar
z)\;\{\alpha,\beta\}&=&\;\;\int_{\partial\sigma}dz\; \beta \partial \alpha
+\int_{\partial\sigma}d\bar z\; \beta \bar \partial \alpha\cr & &\cr  &=&-\int_{\partial\sigma}dz\; \alpha \partial \beta
-\int_{\partial\sigma}d\bar z\; \alpha \bar \partial \beta\label{stkslw}\;, \eeqa
where $\partial\sigma$ is the boundary of $\sigma$.

In writing down  scalar field theory  we shall choose the conformal
gauge.  In that case, the free action for a massless scalar field $\phi$ is
\be S_{\phi}^0 = i \int    dz\wedge d\bar z\; \partial\phi\bar\partial\phi\;,\label{frsclractn}
\ee which can then be re-expressed in terms of  
Poisson brackets 
\be S_{\phi}^0 = 2\pi \int   d\mu_0(z,\bar
z)\;\theta_0(z,\bar z)^{-1}\{z,\phi\}\{\bar z,\phi\} \label{fsfactn}
\ee

 It is not necessary to choose a gauge
restricting the metic tensor in the case of gauge theories.  For this introduce a potential one form ${\tt a}=dz\;
{a}+d\bar z\;\bar{a}$ on $P_0$ which gauge transforms as 
\be {\tt a} \rightarrow {\tt a} + d \lambda\;\label{gtcomth}\ee
  The invariant field strength two-form is 
\be {\tt f}=  i{f}\; dz\wedge d\bar z= (\bar\partial {a}-\partial\bar {a})\;dz\wedge d\bar z\label{cmtvfstrh}\ee
Using (\ref{stkslw}) the flux $\Phi^0_\sigma$ through any region
$\sigma$  can be expressed as an integral of
Poisson brackets of ${a}$ and $\bar {a}$
\beqa \Phi^0_\sigma&=& \int_\sigma {\tt
  f}\;=\;\int_{\partial \sigma}{\tt a}\cr & &\cr  &=&2 \pi\;\int_\sigma d\mu_0(z,\bar
z)\;(\{z,{a}\}+\{\bar z,\bar{a}\})\;,\label{stksthm}\eeqa  having no dependence on the metric, since $\theta_0$ appearing in the
measure cancels with  $\theta_0$ appearing in the Poisson brackets.  In
two dimensions the standard quadratic field action  depends on the metric only through its
determinant, and like the flux,  its integrand can be expressed solely in terms
of the
Poisson brackets of ${a}$ and $\bar {a}$
\beqa S_f^0&=& \int_\sigma \frac{d^2x}{\sqrt{g}}\; f^2 =\frac
i4\int_\sigma  dz\wedge d\bar z \;\theta_0(z,\bar z)\;{f}^2\\ & &\cr &=& \frac\pi2 \int_\sigma d\mu_0(z,\bar
z)\;(\{z,{a}\}+\{\bar z,\bar{a}\})^2\label{loactn}\eeqa  
In comparing with the  free scalar field action (\ref{fsfactn}), here we have not specified
a  gauge for the metric and  $\theta_0$ does not appear explicitly in the
integrand, despite its implicit dependence.
  There are no propagating degrees of freedom for this
system.  Rather, the equations of motion imply that 
\be f=\frac {C_0}{\theta_0(z,\bar z)}\;,\label{slncl}\ee
where $ C_0$ is the  constant associated with  the flux per unit area
\be C_0= \frac{ \Phi^0_\sigma}{2\pi \int_\sigma d\mu_0(z,\bar z)}
\label{fpua}\ee
The action per unit area of the solution is also a constant, namely $C_0^2/4$.

\section{Flat  space - Noncommutative theory}
\setcounter{equation}{0}

Now we review field theory on the noncommutative plane.   This is the
case of constant non-commutativity and no curvature.  The algebra is
generated by the operator  ${\bf z}$ and its hermitean conjugate 
${\bf z}^\dagger$, satisfying
\be [ {\bf
  z},{\bf z}^\dagger ] =\kbar \;,\label{cnstthta}\ee  where  $\kbar$ denotes the non-commutativity
parameter.  It is standardly realized  using the Groenewold-Moyal star product
$\star_M$
\be\star_M\;\;=\;\;
\exp{\;\frac{\kbar}2 \;\biggl(\overleftarrow{ \frac\partial{ \partial z }}\;\;
\overrightarrow{ {\partial\over{ \partial\bar z} }}\;-\;
\overleftarrow{ \frac\partial{ \partial\bar z }}\;\;
\overrightarrow{ {\partial\over{ \partial z}
  }}\biggr)}\;,\label{Mylstr}\ee where  the complex coordinates  $z$
and $\bar z$ are now  symbols of   ${\bf z}$ and  
${\bf z}^\dagger$, respectively.
 Upon defining the star commutator of any two
functions ${\cal A}$ and ${\cal B}$ on the complex plane  according to
$[{\cal A},{\cal B}]_{\star_M} \equiv  {\cal A}\;\star_M\;{\cal B} -
{\cal B}\;\star_M\; {\cal A}$ one has, for example, $ [z,\bar
z]_{\star_M}=\kbar\;$.  The standard convention for the integration measure is
\be d\mu_M(z,\bar z) = \; \frac i{2\pi\kbar}\; dz\wedge d\bar z \;\label{gwmsr}
 \ee
 $\kbar$ has units of length-squared and hence the measure  is
  dimensionless, unlike the commutative measure in (\ref{clsclmsr}).  A well known  identity results from  the  star product
$\star_M$
\be \int d\mu_M(z,\bar z)\; {\cal A}\; \star_M\; {\cal B}  =\int
d\mu_M(z,\bar z)\; {\cal A}  {\cal B}\;, \label{idfrmylstr} \ee where
${\cal A}$ and $  {\cal B}$ vanish sufficiently rapidly at infinity, and from
this,  the cyclic property of trace easily follows.

The free scalar field action on the noncommutative plane is well known
\beqa   S_{\phi}^M& =&-\frac {2\pi}\kbar \int   d\mu_M(z,\bar
z)\;[z,\phi]_{\star_M}\; \star_M\; [\bar z,\phi]_{\star_M}\cr & &\cr &
=&-\frac {i}{\kbar^2} \int   dz\wedge d\bar z\;[z,\phi]_{\star_M} [\bar z,\phi]_{\star_M}\;, \label{fsfactnmp}
\eeqa
and from  the fact that derivatives in $z$ and
$\bar z$ are realized
by $\partial =-\frac 1\kbar [\bar
z,\;]_{\star_M}  $ and $\bar\partial =\frac 1\kbar [
z,\;]_{\star_M}  $, respectively,  (\ref{fsfactnmp}) is identical to the free scalar field action on the commutative plane.

For gauge theories on the noncommutative plane we replace the  the
potential one form ${\tt a}$ by $dz
{\cal A}+d\bar z\bar{\cal A}$.  Infinitesimal gauge variations
by $\Lambda$ of  ${\cal A}$ and $\bar
{\cal A}$ are given by
\beqa \delta {\cal A}& =&   -\frac 1\kbar [\bar
z,\Lambda]_{\star_M}  -i[{\cal A},\Lambda]_{\star_M}\cr & &\cr\delta  \bar {\cal
  A}& =& \;\; \frac 1\kbar [
z,\Lambda]_{\star_M}    -i[\bar {\cal
  A},\Lambda]_{\star_M}\;\label{gtmylwl}\eeqa   The field strength two-form is $ i{\cal F}_M\;
dz\wedge d\bar z$, where
\be i{\cal F}_M= \frac 1\kbar [
z, {\cal A}]_{\star_M}+\frac 1\kbar [\bar
z,\bar {\cal A}]_{\star_M} +i[{\cal
  A},\bar {\cal A}]_{\star_M} \;,\label{fsmlwl}\ee which transforms covariantly
  under  gauge transformations,
\be \delta {\cal F}_M=-i[{\cal
  F}_M,\Lambda]_{\star_M}\label{gtrofsmw} \ee   ${\cal F}_M$ can be
also be expressed as
\be {\cal F}_M= [{\cal
  Z},\bar {\cal Z}]_{\star_M}+\frac 1{\kbar^2}[z,\bar z]_{\star_M}
\;\label{fintrmsxxb}\;,\ee where
\be {\cal Z }= -\frac i\kbar \bar z + {\cal A}\qquad \bar {\cal Z}= \frac i\kbar  z
+\bar {\cal A}\label{dfofclXbX}\;, \ee which also transform
covariantly,
\be \delta {\cal Z} =    -i[{\cal Z},\Lambda]_{\star_M}\qquad\delta
\bar {\cal Z} =     -i[\bar {\cal Z},\Lambda]_{\star_M}\;,\label{chigtmylwl}\ee
The standard gauge theory action on the noncommutative plane is
\beqa S_f^M&=& \frac{\pi\kbar}{2} \int d\mu_M(z,\bar z)\;{\cal
  F}_M\;\star_M\;{\cal
  F}_M\cr & &\cr &=&\frac {i}4\int dz\wedge d\bar z\;{\cal
  F}_M^2\;\label{atnnncpln}\eeqa
  When
$\kbar\rightarrow 0$,
the star commutator goes to $i\kbar$ times the Poisson bracket (\ref{pbcsct}), with
$\theta_0(z,\bar z)$ equal to one, and so  (\ref{atnnncpln}) reduces
to $S^0_f$ with the flat metric.
  The free field equations following from variations of
${\cal A}$ and $\bar{\cal A}$ are 
\be [{\cal Z}, {\cal F}_M]_{\star_M} = [\bar {\cal Z}, {\cal F}_M]_{\star_M} =
0\; \ee
They  are solved for ${\cal F}_M  $ proportional to the identity.   For the case of a pure
gauge solutions
(${\cal F}_M = 0$), ${\cal Z}$ and $\bar{\cal Z}$ are given by
\beqa { \cal Z}_{pg} &=&-\frac i\kbar\;\bar U \;\star_M\; \bar
z\;\star_M\; U \cr & &\cr
\bar {\cal Z}_{pg}&=&\;\; \frac i\kbar\; \bar U\; \star_M \; z\;\star_M \;U \;, \label{prgg}\eeqa where $U$
are unitary functions on the complex plane with regard to the  Groenewold-Moyal star product,
$\bar U\; \star_M\; U = U\;\star_M\; \bar U = 1$.  

It is straightforward to couple the  scalar field to gauge theories
on the noncommutative plane.  For this replace (\ref{fsfactnmp}) by
\beqa   S_{\phi,{\cal Z}}^M &=&-{2\pi}\kbar \int   d\mu_M(z,\bar
z)\;[\bar {\cal Z},\phi]_{\star_M}\; \star_M\; [{\cal
  Z},\phi]_{\star_M}\cr & &\cr&=& -i \int   dz\wedge d\bar z\;[\bar {\cal Z},\phi]_{\star_M} [{\cal
  Z},\phi]_{\star_M}  \label{gdfsfactnmp}
\eeqa  Gauge invariance follows from (\ref{fintrmsxxb}) and $ \delta
{\phi} =    -i[{\phi},\Lambda]_{\star_M}$.  The coupled field equations
resulting from variations of the combined action $ S_f^M+
S_{\phi,{\cal Z}}^M$ are
\beqa [{\cal Z}, {\cal F}_M]_{\star_M}+ 2 [\phi,[{\cal
  Z},\phi]_{\star_M}]_{\star_M} & =& 0\cr & &\cr [\bar {\cal Z}, {\cal
  F}_M]_{\star_M}- 2 [\phi,[\bar {\cal Z},\phi]_{\star_M}]_{\star_M} &=&
0\;\cr & &\cr
 [{\cal Z},[\bar {\cal Z},\phi]_{\star_M}]_{\star_M}+ [\bar{\cal Z},[ {\cal Z},\phi]_{\star_M}]_{\star_M}&=&0\label{sclrggfes} \eeqa

\section{Curved space -  Noncommutative theory}
\setcounter{equation}{0}

In  the previous section,  field theories on the noncommutative plane
can be expressed in
terms of commuting  inner  derivatives $\partial =-\frac 1\kbar [\bar
z,\;]_{\star_M}  $ and $\bar\partial =\frac 1\kbar [
z,\;]_{\star_M}  $, satisfying the usual Leibniz rule.  This, however,  is 
not in general possible for non-constant non-commutativity.   Fortunately,  two dimensional
noncommutative field theories can be expressed purely algebraically,
without relying on the notion of
derivatives.  Non-constant non-commutativity in two dimensions means we replace
(\ref{cnstthta}) by \be [ {\bf
  z},{\bf z}^\dagger ] =\Theta( {\bf
  z},{\bf z}^\dagger) \; \label{nclgbr}\ee  for some function 
$\Theta( {\bf  z},{\bf z}^\dagger)$.  (\ref{nclgbr}) defines an
algebra associated with some noncommutative manifold ${\cal M}$. In addition to being a function
of the generators  ${\bf
  z}$ and   ${\bf z}^\dagger$ of the algebra, $\Theta$ depends on an
additional parameter, the non-commutativity parameter, which we again denote by $\kbar$. 
  The
Groenewold-Moyal star product is not very convenient to realize this
algebra since then  $z$ and $\bar z$ appearing in its definition (\ref{Mylstr}) cannot be symbols of ${\bf
  z}$ and   ${\bf z}^\dagger$.  A more convenient associative  star product was
developed in \cite{Alexanian:2000uz} which has an exact integral
expression, and  will be reviewed in
subsection 6.1.  Here we denote it by $\star$, and so if
 $z$ and $\bar z$  $\in{\mathbb{C}}$ are symbols of ${\bf
  z}$ and   ${\bf z}^\dagger$, respectively, then
\be [z,\bar z]_{\star} \equiv  z\star \bar z -
\bar z\star z =\theta(z,\bar z) \;,\label{fndmlstrcmtr}\ee where $\theta(z,\bar z)$
denotes the symbol of $\Theta( {\bf
  z},{\bf z}^\dagger)$.   Now in general we won't have the analogue of the identity (\ref{idfrmylstr}).
 On the other hand, the appropriate integration measure $d\mu(z,\bar z)$ will be
 required to satisfy \be \int d\mu(z,\bar z)\;[{\cal A},{\cal B}]_{\star}=0
\label{csoftr}\;\ee   for any
 functions ${\cal A}$ and
 ${\cal B}$ that fall off sufficiently rapidly at infinity. (\ref
 {csoftr}) corresponds to the cyclic
 property of the trace.  In the  subsection 6.2 we shall use this property
 and the definition of the $\star$ to perturbatively
 construct the measure.   We assume that like the measure $
 d\mu_M(z,\bar z)$ on the noncommutative plane,  $d\mu(z,\bar z)$ is dimensionless.
  
To recover the systems of the previous two sections we will need to examine two limits:

\noindent {i)}  {\it The commutative limit.}  This is
$\kbar\rightarrow 0$.      
   We assume that $\Theta( {\bf  z},{\bf z}^\dagger)$, and consequently $\theta(z,\bar z)$, is linear in $\kbar$ in this limit,
\be \theta(z,\bar z)\rightarrow \kbar\; \theta_0 (z,\bar
z)\label{lwstrdrtht}\;, \ee where $ \theta_0 (z,\bar
z)$  is a dimensionless function which is  independent of $\kbar$.  We
   shall identify it with  $ \theta_0 (z,\bar
z)$ appearing in (\ref{clsclmsr}) and (\ref{pbcsct}), and require that
   the star product goes to the point-wise product, while 
the star commutator goes to $i$ times the Poisson bracket in this
limit. The  commutative limit of the measure $d\mu(z,\bar
 z)$ is  $d\mu_0(z,\bar
 z)/\kbar$,  where $d\mu_0(z,\bar
 z)$ was given in
 (\ref{clsclmsr}) ( $\kbar$ is introduced since
 $d\mu_0(z,\bar z)$ has units of length-squared).
 
\noindent ii)  {\it  The noncommutative plane limit.}  This is  \be\theta({ z},\bar{ z})
\rightarrow\kbar \ee  
The combination of both of the above limits gives the commutative plane.

Ordering ambiguities occur in deforming the free scalar actions
(\ref{frsclractn}) and  (\ref{fsfactnmp}) of the previous two
sections to this general case.  Moreover, here we need  that
$\Theta( {\bf
  z},{\bf z}^\dagger)$ is nonsingular.  We can choose the ordering
such that the general massless scalar field action is
\be  S_{\phi} =- {2\pi} \int   d\mu(z,\bar
z)\;\theta(z,\bar z)^{-1}_\star\star[z,\phi]_\star \star [\bar
z,\phi]_\star \label{frsclractngc}\;, \ee where $\theta(z,\bar
z)^{-1}_\star$ is defined by  $\theta(z,\bar
z)^{-1}_\star\star\theta(z,\bar z)= \theta(z,\bar z)\star\theta(z,\bar
z)^{-1}_\star=1$.
One  easily recovers 
(\ref{frsclractn}) from (\ref{frsclractngc}) in limit i), and
(\ref{fsfactnmp}) in limit ii).  The field equation following from
variations of $\phi$ in (\ref{frsclractngc}) read
\be [\;[\bar
z,\phi]_\star\star\theta^{-1}_\star\;,\;z\;]_\star+[\;\theta^{-1}_\star\star[z,\phi]_\star\;,\;\bar
z\;]_\star = 0\label{gfefsfnc} \ee
It has the trivial solutions $\phi=z$ and $\phi=\bar z$, as well as the
constant solution,  but the analytic
and anti-analytic solutions  of the commutative theory,
$\phi=\phi_+(z)$ and $\phi=\phi_-(\bar z)$, are not in general
present in the
noncommutative theory.

The situation  is more straightforward for gauge theories.  Since the
commutative action  (\ref{loactn}) could be expressed without explicit
reference to $\theta_0$ the above ordering ambiguity does not arise,
and, moreover, no space-time gauge was necessary in writing down (\ref{loactn}).   Upon again introducing potentials ${\cal A}$ and $\bar
{\cal A}$ we can  define the field strength by
 \be i {\cal F}=\frac 1\kbar[{ z},{ \cal A}]_\star+\frac 1\kbar[ {\bar z} ,{\bar A}]_\star+i[{
  {\cal A}},\bar{\cal A}]_\star
\label{smblsfs}\;,\ee and generalize 
 the commutative flux (\ref{stksthm}) in some
 region $\sigma$ on the complex plane to
 \beqa
 \Phi_\sigma &=& 2\pi\kbar\int_\sigma  d\mu(z,\bar z)\;{\cal F}
\label{gnrlflx}\;\eeqa
  From (\ref{csoftr}), it is zero
   if $ {\cal A}$ and $\bar {\cal A}$ vanish on the boundary
   $\partial\sigma$ of $\sigma$. 
 The  action (\ref{atnnncpln}) on the noncommutative plane can be
   generalized  to \beqa
 S_f &=& \frac{\pi\kbar}2\; \int d\mu(z,\bar z)\;{\cal F}\star{\cal
   F}\;
\label{ncactnsmbl}\eeqa  In the commutative limit i), 
${\cal F} \rightarrow \theta_0 f$, and (\ref{ncactnsmbl}) reduces to
the commutative Maxwell action (\ref{loactn}).
In the noncommutative plane limit ii), the field strength (\ref{smblsfs}) becomes
  (\ref{fsmlwl}), and we  recover (\ref{atnnncpln}) from
(\ref{ncactnsmbl}).  
The noncommutative Maxwell equations following from variations of
${\cal A}$ and $\bar{\cal A}$ in (\ref{ncactnsmbl}) (ignoring boundary
terms) are 
\be [{\cal Z}, {\cal F}]_\star = [\bar {\cal Z}, {\cal F}]_\star =
0\;, \label{eomfmes}\ee
where ${\cal Z}$ and $\bar{\cal Z}$ were defined in
(\ref{dfofclXbX}).  They are again  solved  for ${\cal F}  $
proportional to the identity.   Then from (\ref{gnrlflx}) and
(\ref{ncactnsmbl}), respectively,
the flux per unit area and action per unit area for any solutions  are
constants, just as in the commutative case.  Now the area
of any region $\sigma$ on the complex plane  is  given by $\int_\sigma d\mu(z,\bar z)$.   We have
not found a simple expression for pure gauge solutions, analogous
to those on the noncommutative plane  (\ref{prgg}), although an
expansion about the commutative answer can be obtained.  We do this in
subsection 6.4.

 The issue of gauge invariance of  the action
(\ref{ncactnsmbl}) is more complicated than it was for the previous two limits.
Applying  (\ref{csoftr}), gauge invariance of the action follows  if the field strength transforms
covariantly, i.e.  variations are of the form 
\be \delta {\cal F}= -i[{\cal
  F},\Lambda]_{\star}\;,\label{gtrofsgc} \ee for infinitesimal
$\Lambda$. 
Although the field strength transforms in a simple manner, the same
 is not, in general, true for the potentials ${ \cal A}$ and
 ${\bar {\cal A}}$.  For example, something along the lines of
 (\ref{gtmylwl}) does not work because $\theta$ does not have zero
 star commutator with $\Lambda$.  The gauge symmetry of the action is therefore
 hidden.   Here it does not help to introduce  the
 quantities ${\cal Z}$ and $\bar{\cal Z}$ defined in (\ref{dfofclXbX})
 and express  ${\cal F}$
 according to \be  {\cal F}=[{\cal
  Z},\bar {\cal Z}]_{\star}+\frac 1{\kbar^2}\theta(z,\bar z) \;,\label{fsitfzbz} \ee in
analogy to  (\ref{fintrmsxxb}).    
 Since $\theta(z,\bar z)$ is not  covariant, neither can be ${\cal Z}$
 and $\bar{\cal Z}$.  In  subsection 6.4 we develop a
 perturbation scheme for determining how  ${ \cal A}$ and
 ${\bar {\cal A}}$, or equivalently  ${\cal Z}$
 and $\bar{\cal Z}$, gauge transform away from limits i) and ii).

The  scalar field action coupled to gauge theories on the
noncommutative plane (\ref{gdfsfactnmp})
can be generalized to arbitrary $\theta(z,\bar z)$ 
 by
\be   S_{\phi,{\cal Z}} =-{2\pi}\kbar^2 \int   d\mu(z,\bar
z)\;\theta(z,\bar z)^{-1}_\star\star [\bar {\cal Z},\phi]_{\star}\; \star\; [{\cal Z},\phi]_{\star} \label{gdfsfactnmpgtht}
\ee   For gauge invariance we need that \beqa \delta [{\cal Z},
{\phi}]_\star &=&    -\frac i\kbar\; [[{\cal Z}, {\phi}]_\star,\Lambda]_{\star}\star
\theta \cr & & \cr  \delta [\bar {\cal Z},
{\phi}]_\star &=&    -\frac i\kbar\;\theta \star[[\bar {\cal Z},
{\phi}]_\star,\Lambda]_{\star}\label{chiphigtmylwl}\eeqa  It is then
clear that the scalar field  $\phi$ cannot, in general, transform
covariantly.  After developing a
 perturbation scheme for the gauge transformation of the potentials, one  can then  use (\ref{chiphigtmylwl}) to do the
 same for $\phi$. The coupled fields equations (\ref{sclrggfes}) are
 then generalized to
\beqa [{\cal Z}, {\cal F}]_{\star}+ 2\kbar\; [\;\phi\;,\;[{\cal
  Z},\phi]_{\star}\star\theta^{-1}_\star\; ]_{\star} & =& 0\cr & &\cr [\bar {\cal Z}, {\cal
  F}]_{\star}- 2\kbar\;  [\;\phi\;,\;\theta^{-1}_\star \star [\bar {\cal Z},\phi]_{\star}\;]_{\star} &=&
0\;\cr & &\cr [\;[{\cal
  Z},\phi]_\star\star\theta^{-1}_\star\;,\;\bar{\cal
  Z}\;]_\star+[\;\theta^{-1}_\star\star[\bar{\cal
  Z},\phi]_\star\;,\;{\cal Z}\;]_\star &=&0 \eeqa

\section{Magnetic Monopoles on the Fuzzy sphere}

\setcounter{equation}{0} 

Fuzzy spaces are standardly defined to be noncommutative theories with
finite dimensional matrix representations.  So in that case the generators $ {\bf
  z}$ and ${\bf z}^\dagger$ of the algebra  in (\ref{nclgbr})   are represented 
$N\times N$ matrices.  This also applies to  the fields $\phi$, $
{\cal Z}$ and $\bar {\cal Z}$  which are polynomials functions of  $ {\bf
  z}$ and ${\bf z}^\dagger$.  The star can be replaced by ordinary matrix
multiplication, and so the star commutator is replaced by the matrix
commutator. Integration corresponds to taking the trace.  Specializing
to gauge theories, one gets that the  field strength is traceless and the total flux
vanishes Tr$\;{\cal F}=0$.  Furthermore, the  constant solution,
i.e. ${\cal F} $ proportional to the identity,  to the noncommutative Maxwell
equations (\ref{eomfmes}) collapses to the trivial solution, i.e. ${\cal F}=0$.   This indicates the absence of any
magnetic monopole solutions  in a fuzzy physics.  In deriving
(\ref{eomfmes}) one assumed arbitrary variations of
the gauge fields in the Maxwell action (\ref{ncactnsmbl}).  The
negative result for monopoles can be avoided if we restrict variations
of $ {\cal Z}$
and $\bar {\cal Z}$ to  block diagonal matrices.  In that case
${\cal F}$ has solutions which are block diagonal  matrices, with
the individual blocks being proportional to
 identity matrices, and their combined trace equal to
 zero.   We shall use this technique to construct fuzzy magnetic monopole
solutions in subsection 5.2.\footnote{Although the procedures differ, reducible representations were also necessary for
  describing   monopoles in \cite{Karabali:2001te}.}
   In subsection 5.1  we  review the commutative case.

\subsection{Commutative case}
  
We shall fix the radius of the sphere to be $1$, so in terms of  embedding coordinates 
$x_i,\;i=1,2,3$, $ x_1^2+x_2^2+x_3^2 = 1$.  Poisson brackets which
preserve the $SO(3)$ symmetry are
\be \{x_i,x_j\}=\epsilon_{ijk}x_k\;\ee  In defining gauge theory, one
can introduce 3-potentials $a_i,\;i=1,2,3$,
which gauge transform like\cite{Jayewardena:1988td}
\be a_i\rightarrow a_i +\{x_i,\lambda\} \;,\label{gtic}\ee for some function
$\lambda$ on the sphere.  A  constraint should be imposed on $a_i$ as there are  only two independent
gauge potentials on the surface.  It should not restrict the gauge
transformations (otherwise it would be a gauge condition), and  be invariant
under rotations.  This  is the case for
\be a_ix_i =0 \label{ccgp}\ee     From this one gets the identity
\be x_j\{x_i,a_j\}=\epsilon_{ijk} x_j a_k \;,\label{id1fmadx}  \ee
in addition to \be x_i\{x_i,a_j\}=0  \label{id2fmadx} \ee 
  A gauge invariant scalar is
\be b= -\epsilon_{ijk}\; x_i \{x_j, a_k\}\label{mfiec}\;, \ee which
can be interpreted as  the magnetic flux density normal to the surface.
The
Maxwell action on the sphere is \be S_f^0=\frac1{4\pi} \int d\Omega \; b^2
\label{actnnmbdngcrd}\;,\ee where the integral is over the solid angle $ \Omega$.

To recover the formalism of sec. 2, we can
 stereographically  project to the complex plane, where the
north pole is mapped to infinity thus corresponding to a coordinate singularity
\be z =\frac{x_1 -ix_2}{1-x_3}\qquad\bar z =\frac{x_1
  +ix_2}{1-x_3}\;\ee   The Poisson structure is then projected to
\be \{z,\bar z\} =-
i\theta_0(|z|^2)\;,\qquad\theta_0(|z|^2)=\frac 12 (1+|z|^2)^2
\;,\label{sphprjtdpb} \ee  while the potentials $a_i$ are mapped to the one form ${\tt a}=dz\;
{a}+d\bar z\;\bar{a}$, where
\be 2i{a} = (1-x_3)(a_1+i a_2)+ (x_1+ix_2) a_3\;,\label{twoia} \ee and
 $\bar a$ is the complex conjugate of $a$.
The inverse map is  
\beqa  a_1+ia_2 &=& i({a} + \bar z^2\bar{a}) \cr
 a_1-ia_2 &=&- i(\bar{a} + z^2{a}) \cr
a_3&=& i(z {a} -\bar z \bar {a})\eeqa
From  (\ref{gtic}) one recovers 
 the gauge transformations (\ref{gtcomth}).
  The magnetic flux
density (\ref{mfiec})   is mapped to\footnote{ To prove this
 substitute  (\ref{twoia}) in (\ref{dfnfmfns})  to get $(1-x_3)( \{z,{a}\} +
 \{\bar{z},\bar a\})= b + \epsilon_{ij3}\{x_i,a_j\} -a_3 $ and use the
 identity $x_3 b= a_3-\epsilon_{ij3}\{x_i,a_j\}$, which follows from 
(\ref{id1fmadx}) and (\ref{id2fmadx}).}  \be {b} = \{z,{a}\} +
 \{\bar{z},\bar a\}\label{dfnfmfns} \ee  Thus, by doing the
 sterographic projection of the Maxwell action (\ref{actnnmbdngcrd}) we
 are able to recover the expression (\ref{loactn}).

  The magnetic
 monopole solutions are $b=C_0= \frac{g_0}{4\pi}$, or
\be f= \frac {g_0}{2\pi} \;\frac 1{(1+|z|^2)^2}\;,
 \label{mmcstslnlp}\ee
where $g_0$ is  the magnetic charge.  The Maxwell action
(\ref{actnnmbdngcrd}) evaluated for this solution is $  
(\frac{g_0}{4\pi})^2$.
  Potentials can be given after
removing the point at infinity, the location of the Dirac string, 
\be {\tt a} = \frac{ig_0}{4\pi}\; \frac {{\bar z}dz - z d{\bar z}}
{1+|z|^2}\; \label{cmnplptl} \ee

\subsection{Noncommutative case}
In going to the fuzzy sphere, we replace real coordinates $x_i$  by hermitean operators ${\bf x}_i$,
 satisfying commutation relations:
\be [{\bf x}_i,{\bf x}_j] = {i \kbar}\;\epsilon_{ijk}{\bf x}_k\;,\label{xixj}\ee
as well as   $ {\bf x}_i{\bf x}_i=\BI$, $\BI$ being the unit operator.  When the non-commutativity  parameter $ \kbar$ has values $\kbar_J={1/ {
\sqrt{J(J+1)}}}\;,$ $ J={1\over 2}, 1, {3\over 2},...\;,$  ${\bf x}_i$
 have finite dimensional representations.   For a given $J$, one can set ${\bf x}_i=\kbar_J {\bf J}_i$,
 ${\bf J}_i$ being the angular momentum matrices associated with   
 the $2J+1$ dimensional irreducible representation  $\Gamma^J$ acting on
 Hilbert space $H^J$ with  states $|J,M>$, $M=-J,-J+1,...,J$.   The commutative
limit is $J\rightarrow\infty$ corresponding to infinite dimensional representations. 

A nonsingular fuzzy stereographic projection was given in
\cite{Alexanian:2000uz}.\footnote{A singular one was given in
\cite{Morariu:2005pv}.}   It is defined up to an operator ordering ambiguity to be 
\be {\bf z} =({\bf x}_1-i {\bf x}_2)(1-{\bf x}_3 )^{-1} \;,\qquad
   {\bf z}^\dagger =(1-{\bf x}_3 )^{-1} ({\bf x}_1+i {\bf x}_2)
   \label{fzysp} \ee  This is a nonsingular map because $1$ is excluded from
   the spectrum of ${\bf x}_3$, except in the commutative limit
   $J\rightarrow\infty$.  For the top state $M=J$, ${\bf x}_3$  has eigenvalue
   $J/\sqrt{J(J+1)}$, which approaches $1$ in the limit, and we
   thereby recover the
   coordinate singularity.  Using properties of angular momentum
   matrices,  $\Theta( {\bf
  z},{\bf z}^\dagger)$ is represented by a diagonal matrix
\be \Theta( {\bf
  z},{\bf z}^\dagger)|J,M> =\theta_{J,M}|J,M> \;,\ee  with elements
\be \theta_{J,M}\;=\;\frac{J(J+1) - M(M+1)}{(\sqrt{J(J+1)} - M -1)^2}\;-\;\frac{J(J+1) - M(M-1)}{(\sqrt{J(J+1)} - M )^2}\;.\label{lajm}\ee
When evaluated on the top state one has $\theta_{J,J}=
   -2J/(\sqrt{J(J+1)} - J )^2$, which goes like $-8J$  in the limit
   $J\rightarrow\infty$.

In constructing gauge theories, as stated previously, the field strength expressed in terms of potentials,
   (\ref{smblsfs}) or
   (\ref{fsitfzbz}), is traceless, and as a
   result  the total flux
vanishes, i.e.  Tr$\;{\cal F}=0$.   This  implies that there can be no magnetic monopoles in
   this formalism, and furthermore that the constant solution  to the noncommutative Maxwell
equations (\ref{eomfmes})
   collapses to the trivial solution, i.e. ${\cal F}=0$.  This is not surprising since also in the
   commutative theory, if we insist on
   writing the field strength in terms of potentials globally,  there
   can be no magnetic monopoles.  The
   monopole potential in (\ref{cmnplptl})  is defined only after
removing the point at infinity from the domain  of the commutative
   theory.  In the noncommutative theory, this point is
   approached by  the top state as $J\rightarrow \infty$. So let us
   similarly remove it from the 
   domain of the noncommutative theory.  Equivalently, we can restrict
 {\it   variations}  of
   the fields  $ {\cal Z}$
and $\bar {\cal Z}$ to be block diagonal matrices, one block being
   $2J\times 2J$ and the other being $1\times 1$,  the latter
   associated with  the top state.  Then the equations of motion
   (\ref{eomfmes}) will only hold for the diagonal blocks.
Solutions to the noncommutative Maxwell
equations (\ref{eomfmes}) for ${\cal F}$ will then also be be block diagonal
   matrices ${\cal F}_{2J\times 2J}$ and  ${\cal F}_{1\times 1}$,
   which are proportional to the
 identity,  and since Tr$\;{\cal F}=0$,
\beqa {\cal F}_{2J\times 2J}&=& \frac
   g{2J}\;\BI_{2J\times 2J}\\ & &\cr  {\cal
   F}_{1\times 1}&=&- g\;,\label{ncmnplsns} \eeqa $\BI_{2J\times 2J}$ being
   the  $2J\times 2J$ identity matrix. The trace over only $ {\cal
   F}_{2J\times 2J}$ is $g$, which by analogy with the commutative theory 
   defines  the magnetic charge, while ${\cal
   F}_{1\times 1}$ corresponds to  the compensating  flux of the Dirac
   string.  So (\ref{ncmnplsns}) is the noncommutative analogue of the
   Dirac string.  The action (\ref{ncactnsmbl}) evaluated for this
   solution gives
 \be  \frac{\pi g^2 (2J+1)}{4J\sqrt{J(J+1)}} \ee
In comparing with the commutative answer of  $  
(\frac{g_0}{4\pi})^2$, we need that $g$ grows like
   $\sqrt{J}$ in the commutative limit, i.e.
  \be g \rightarrow \frac{\sqrt{J}\; g_0}{(2\pi)^{3/2}} \;\;,\qquad {\rm
   as}\;\;J\rightarrow \infty\label{cmtvlmtfrg}\;, \ee in order to
   recover a finite charge $g_0$ in the commutative theory.

 By equating
    (\ref{ncmnplsns}) with the last row and column of the matrix
   associated with 
   (\ref{fsitfzbz}) we get  \be
 g =\sum^{J-1}_{M=-J}\biggr(|<J,M| {\cal Z}|J,J>|^2-| <J,M|\bar  {\cal
   Z}|J,J> |^2 \biggl) +\frac{2J^2(J+1)}{(\sqrt{J(J+1)} - J )^2}
\label{ncmcnq} \ee  Note that only off-block diagonal matrix elements of  $ {\cal Z}$
and $\bar {\cal Z}$ are present in the result.  These are
   non-dynamical fields (they were not varied in obtaining the field
   equations), and so {\it the magnetic
   flux is a constant of the motion.}  This is a result of the
   dynamics, and not topology.   In the commutative limit, the
   last term in (\ref{ncmcnq}) goes to infinite like $8J^3$, and thus
   the
   sum must go like $-8J^3$ in 
   order  for $g$ to have the  limit in (\ref{cmtvlmtfrg}).  Alternatively, we can use
   (\ref{smblsfs}) to write the charge in terms of  matrix elements of $ {\cal A}$
and $\bar {\cal A}$
 \beqa
 g &=&\sum^{J-1}_{M=-J}\biggr(|<J,M| {\cal A}|J,J>|^2-| <J,M|\bar
   {\cal A}|J,J> |^2 \biggl)\cr & &\cr
& &-\;i\; \frac{J\sqrt{2(J+1)}}{\sqrt{J(J+1)}-J}\;\biggl(<J,J| {\cal A}|J,J-1>- <J,J-1|\bar
   {\cal A}|J,J>  \biggr)
\label{ncmcnqita} \eeqa  Again, only the  (non-dynamical)  off-block diagonal matrix
   elements appear, and so $g$ is a constant of the motion. 

In the above, although the  charge is a constant of motion, we
get no quantization, at least at the classical
level.\footnote{Quantized magnetic charges were obtained in
\cite{Karabali:2001te} upon requiring the gauge fields to be
associated with
reducible $SU(2)$ representations.}   On the other hand, quantization
does occur  if we impose the stronger condition
that   the fields  $ {\cal Z}$
and $\bar {\cal Z}$, and not just their variations, are block
diagonal.   
 Then the sum in (\ref{ncmcnq}) vanishes and $g$ is just equal to the
 last term.
 We can also allow for more general block diagonal matrices.  So let    $ {\cal Z}$
and $\bar {\cal Z}$ be reducible to 
  $(2J+1-N)\times (2J+1-N)$ and  $N\times N$ matrices, $1\le N\le 2J$.
Solutions to the noncommutative Maxwell
equations (\ref{eomfmes})  for ${\cal F}$ will then  be block diagonal
   matrices ${\cal F}_{(2J+1-N)\times (2J+1-N)}$ and  ${\cal
     F}_{N\times N}$, where
\beqa {\cal F}_{(2J+1-N)\times (2J+1-N)}&=& \frac
   g{2J+1-N}\;\BI_{(2J+1-N)\times (2J+1-N)}\cr & &\cr  {\cal
   F}_{N\times N}&=&-\frac gN\;\;\BI_{N\times N}\;\label{qncmnplsns}
 \eeqa   So  now the noncommutative analogue of the
   Dirac string includes $N$ states. By equating
   the trace of either of the matrices in (\ref{qncmnplsns})  with the
   corresponding trace  of
   (\ref{fsitfzbz}) we get \be
 g = -\frac 1{\kbar_J^2}\sum^N_{n=1}
   \theta_{J,J+1-n} \ee
Using (\ref{lajm}) we thereby get quantized magnetic charges, depending on $J$ and $N$.
   Examples are
\beqa  J=\frac 12\;,& N=1&\qquad g=\frac3{(\sqrt{3}-1)^2} \cr& &\cr
 J= 1\;,& N=1&\qquad g=\frac4{(\sqrt{2}-1)^2} \cr & &\cr
& N=2&\qquad g=2\cr & &\cr  J=\frac 32\;,& N=1&\qquad
   g=\frac{45}{(\sqrt{15}-3)^2}\cr & &\cr & N=2&\qquad g=\frac{30(8
   \sqrt{15} -31)}{95\sqrt{15}-368}\cr & &\cr & N=3&\qquad g=\frac{45}{(\sqrt{15}+1)^2}
\eeqa
These charges do not have a well defined  limit when $J\rightarrow\infty$.  Here it
does not help that we have an additional parameter $N$.  For $N$ equals one, or close to one, $g$ goes
like $8J^3$ as  $J\rightarrow\infty$, which is  too divergent when
compared to (\ref{cmtvlmtfrg}).  On the other side is $N=2J$, where
$g=\theta_{J,-J}/\kbar^2_J$.  So when $N$ equals $2J$, or close to $2J$, $g$  goes
like $J/2$ as  $J\rightarrow\infty$, which is also too divergent when
compared to (\ref{cmtvlmtfrg}). 
It follows that off-block diagonal matrices must be present for  $ {\cal Z}$
and $\bar {\cal Z}$ in order to recover the usual commutative limit,
even though these matrices are non-dynamical.

\section{Noncommutative corrections}

\setcounter{equation}{0}

 In this section we do   an expansion in $\kbar$ to obtain noncommutative
 corrections to the commutative scalar and gauge field actions.   We
 also obtain
corrections to the corresponding actions on the noncommutative plane.
 For the latter  we  do an
expansion in derivatives of $\theta({ z},\bar{ z})$. 
The two expansions are not independent as is explained below.  We
 also get an expansion for
 gauge transformations of ${ \cal A}$ and ${\bar {\cal A}}$ about the
two limits.  The expression in general depends on the
 star product and $\theta (z,\bar
z)$, which in the commutative limit is related to the determinant of the metric.   So for
 this gauge theory, motion along a
 fibre  depend on the geometry of  the base manifold.
 The noncommutative gauge theory can be Seiberg-Witten\cite{Seiberg:1999vs} (also
see
\cite{Asakawa:1999cu},\cite{Grimstrup:2003rd},\cite{Pinzul:2004tq})   mapped to the
 commutative theory, leading to corrections to the commutative flux
 (\ref{stksthm}) and the Maxwell action
 (\ref{loactn}).   We   then get the corrections to the commutative
 solution 
(\ref{slncl}).  In section 7 we apply the techniques to
 the example of noncommutative AdS${}^2$. There are a number of
 obstacles in using the approximation scheme  developed  here for
 analyzing magnetic monopoles in fuzzy
 gauge theories, which we comment on in section 8.

\subsection{Star product}

We now review the star product in \cite{Alexanian:2000uz} which can be
expressed in terms of the symbols $z$ and $\bar z\in{\mathbb{C}}$ of
operators ${\bf z}$ and ${\bf z}^\dagger$, respectively, and which
easily reproduces the star commutator  (\ref{fndmlstrcmtr}).
It is based on an
overcomplete set of unit vectors $\{|z>\}$ spanning an infinite dimensional
Hilbert space.   The states $|z>$    are, in general, nonlinear deformations of standard coherent states
 on the complex plane.  They diagonalize ${\bf
z}$,\footnote{It is  problematic to construct such
     states for fuzzy manifolds.  Alternatively, for the case of the
     fuzzy sphere we found   a set of states
     $\{|z>\}$ where the difference of ${\bf  z}|z>$ and $z|z>$ was
     proportional to the top state.\cite{Alexanian:2000uz}  As a result, the expressions
     which follow for the star product and measure get modified for
     the fuzzy sphere.}
   \be {\bf  z}|z>=z|z>\; \label{zee}\ee
   The covariant symbols
of  operators $A$, $B$,... are given by  ${\cal A}(z,\bar z)=<z|A|z>$,
${\cal B}(z,\bar z)$ = $<z|B|z>$,... , and their star product by  
$ [ {\cal A} \star {\cal B}](z,\bar z )=<z|AB|z>$. 
  The
 expression for the
 star product  was obtained in
 \cite{Alexanian:2000uz}. Given the symbols  ${\cal A}(z,\bar z)$ and 
${\cal B}(z,\bar z)$, then  
$ [ {\cal A} \star {\cal B}](z,\bar z )$ can be written  as
 \be {\cal A}
(z,\bar z )\;\;\int d\mu (\eta,\bar\eta )\;   \;
:\exp{ \overleftarrow{ \frac\partial{ \partial z }}(\eta-z)  }:\;
\; |
<z |\eta >  |^2\;  :
\exp{  (\bar\eta-\bar z ) \overrightarrow{ \frac\partial {
      \partial\bar z } } }:\;\;  {\cal B}(z,\bar z
 )\;,\label{iefsp}\ee   where $ d\mu (z,\bar z )  $ is the
 appropriate measure
  on the complex plane satisfying  the
partition of unity $\int d \mu(z,\bar z)\;{|z>}{<z|}=\BI$.   The colons in (\ref{iefsp}) denote an ordered
  exponential, with the derivatives  ordered to the right in each
  term in the Taylor expansion of $  \exp{  (\eta-z )
    \overrightarrow{ \frac\partial { \partial z } } } \;,$ and to
  the left in each term in the Taylor expansion of $\exp{
    \overleftarrow{ \frac\partial{ \partial\zeta }}(\eta-z)  }$; i.e.
\beqa  :\exp{  (\bar\eta-\bar z ) \overrightarrow{ \frac\partial {
       \partial\bar z } } }: \;&=& 1 + (\bar\eta-\bar z ) \overrightarrow{ \frac\partial {
       \partial\bar z } } +\frac 12  (\bar\eta-\bar z )^2 \overrightarrow{ \frac{\partial^2} {
       \partial\bar z^2 } } +\cdot\cdot\cdot
\cr & &\cr
:\exp{ \overleftarrow{ \frac\partial{ \partial z }}(\eta-z)
}: \;&=&1 + \overleftarrow{ \frac\partial{ \partial z }}(\eta-z)
+\frac12  \overleftarrow{ \frac{\partial^2}{ \partial z^2 }}(\eta-z)^2
 +\cdot\cdot\cdot\label{eooe} \eeqa  When  $\theta(z,\bar z)=<z|\Theta( {\bf
  z},{\bf z}^\dagger)|z>$ equals
  $\kbar$, $|z>$ are the standard coherent states and the star product
  reduces to the Voros product, which can be transformed to (\ref{Mylstr}).
 A derivative expansion of the above star product was 
 performed  in \cite{Pinzul:2001my}.   There  one obtained the following leading three terms
acting between functions of $z$ and $\bar z $:
\beqa\star& =&  1\;\;+ \;\;\overleftarrow{ \frac\partial{ \partial
    z }}\; \theta(z,\bar z) \;
\overrightarrow{ {\partial\over{ \partial\bar z} }}\cr & &\cr & &
\;\;
+\;\;
\frac14\biggl[\overleftarrow{\frac{\partial^2}{ \partial z^2 }}
    \;\;  \overrightarrow{ \frac\partial {
      \partial\bar z } }\theta(z,\bar z)^2
  \overrightarrow{ \frac\partial {
      \partial\bar z } }\; +\;
\overleftarrow{\frac{\partial}{ \partial z }}
  \theta(z,\bar z)^2\;\overleftarrow{\frac{\partial}{ \partial z }}
   \;\;
  \overrightarrow{ \frac{\partial^2} {
      \partial\bar z^2 } }\biggr]\;\;+\;\;\cdot\cdot\cdot
\label{afsvt}\eeqa
  At lowest order in $\kbar$, we assume that $\theta(  z,{\bar z})$ is
  given by
(\ref{lwstrdrtht}).  Since we
interpret $\kbar $ as the non-commutativity parameter,   then the
 lowest order in the derivative expansion (\ref{afsvt}) gives the
 commutative product,
 and from the first order term, one gets that  the star commutator goes to $i$ times the Poisson bracket.
 Moreover, after  expanding $ \theta(z,\bar z)$ in $\kbar$,
\be \theta(z,\bar z)=\kbar \theta_0(z,\bar z) + \kbar^2 \theta_1(z,\bar z) +\cdot\cdot\cdot \;,\label{xpnsnfthta}\ee the
 derivative expansion of the star can also be regarded as an expansion in
 $\kbar$.
 So one can use (\ref{afsvt}) to expand about the commutative field theory
 or the field theory on the noncommutative plane.

\subsection{Measure}

We next expand the
 integration measure  about its i) commutative limit $d\mu_0(z,\bar
 z)/\kbar$ and ii) the  noncommutative plane limit $ d\mu_M(z,\bar z)$.  For this we
 require that the trace property  (\ref{csoftr}) holds order-by-order  for functions ${\cal A}$ and
 ${\cal B}$ that fall off sufficiently rapidly at infinity.
Using  (\ref{afsvt}) we then find 
\be d\mu(z,\bar z)=\frac i {2\pi}\frac{ dz\wedge d\bar z}{\theta(z,\bar z)}\; \biggl(1
 +\frac 12 \partial\bar\partial\;\theta(z,\bar z)+\cdot\cdot\cdot\biggr)\label{crtdmsr}
 \ee
We can regard this as a derivative expansion, and thus a perturbation
 about the noncommutative plane limit ii).  (\ref{crtdmsr}) can also
 be regarded as an expansion in $\kbar$, and thus a perturbation
 about the commutative limit i).  For the latter,  apply
 (\ref{xpnsnfthta}) to get
\be d\mu(z,\bar z)=\frac{ d\mu_0(z,\bar z)} \kbar \biggl\{ 1 + \kbar
 \biggl (\frac 12  \partial\bar\partial\;\theta_0 - \frac
 {\theta_1}{\theta_0}\biggr) + \cdot\cdot\cdot \biggr\} \ee

More generally, if real functions ${\cal A}$ and
 ${\cal B}$ and their derivatives are nonvanishing on the boundary $\partial \sigma$ of
 some region $\sigma$ then the integral of
 their star commutator can be expressed on the boundary.  The
 generalization of (\ref{stkslw}) gives
\beqa & &\int_\sigma d\mu(z,\bar z)\;[{\cal
 A},{\cal B}]_{\star}\cr & &\cr &=&\frac i{2\pi} \int_\sigma dz \wedge d\bar z\;\biggl\{
\partial{\cal A}\bar\partial{\cal B} +\frac 12
 \partial\bar\partial\theta\partial{\cal A}\bar\partial{\cal B}\cr&
 &\cr& &\qquad\quad +\;\frac
 1{4\theta} \biggl(\partial^2{\cal A} \bar\partial(\theta^2\bar\partial{\cal B})
+ \partial(\theta^2\partial {\cal A})\bar\partial^2{\cal B}\biggr)\;+\;\cdot\cdot\cdot \;\;\; -\;({\cal A}
 \rightleftharpoons {\cal B})\biggr\}  \cr & &\cr&=&\frac i{2\pi}
 \int_\sigma dz \wedge d\bar z\;(\bar \partial {\cal J}_{[{\cal
 A},{\cal B}]} -  \partial\bar {\cal J}_{[{\cal
 A},{\cal B}]})\cr & &\cr & =&\frac i{2\pi} \int_{\partial \sigma}( dz \; {\cal J}_{[{\cal
 A},{\cal B}]} + d\bar z \; \bar {\cal J}_{[{\cal
 A},{\cal B}]}) \;,\label{twntefv}
  \eeqa with 
\beqa  {\cal J}_{[{\cal
 A},{\cal B}]}&=&\frac12\; {\cal B}\partial {\cal A}\;+\;\frac 14 \biggl(\partial  {\cal B}(\theta \partial\bar\partial {\cal A}-\partial \theta
\bar \partial {\cal A}) - \theta
 \partial^2 {\cal B}\bar \partial{\cal A}\biggr)\;+\;\cdot\cdot\cdot \;\; -\;({\cal A}
 \rightleftharpoons {\cal B})
 \cr & &\cr  \bar {\cal J}_{[{\cal
 A},{\cal B}]}&=&\frac12\; {\cal B}\bar\partial {\cal A}\;+\;\frac 14 \biggl(\bar\partial  {\cal B}(\theta \partial\bar\partial {\cal A}-\bar\partial \theta
 \partial {\cal A}) - \theta
 \bar\partial^2 {\cal B} \partial{\cal A}\biggr)\;+\;\cdot\cdot\cdot \;\; -\;({\cal A}
 \rightleftharpoons {\cal B})\label{twntesxv} \eeqa

\subsection{Field theory}

We can now apply the previous  expansions 
about  limit i) and ii) to the scalar field and gauge theory actions.
Starting with the real scalar field, we get
\beqa  [z,\phi]_\star &=&\;\;\theta\;\biggr( \bar\partial \phi + \frac 12
\partial\theta\bar\partial^2\phi+ \cdot\cdot\cdot\biggr)\cr& &\cr
 [\bar z,\phi]_\star &=&-\theta\;\biggr(\partial \phi + \frac 12\bar
\partial\theta \partial^2\phi+ \cdot\cdot\cdot\biggr) \;\eeqa
After substituting into (\ref{frsclractngc}), and dropping
boundary terms, we can write $S_{\phi}$,  up to first order, by
simply replacing the ordinary derivatives in the commutative action
(\ref{frsclractn}) with `covariant' ones
\beqa S_{\phi}&=& i\int  dz\wedge d\bar z\;
 \;{\cal D}_\theta\phi\;\bar{\cal D}_\theta\phi\;,\label{frdrcrtdactnsc}
\eeqa  where ${\cal D}_\theta$ and $\bar{\cal D}_\theta$ are defined
by\footnote{The geometric
  meaning of these derivatives is not immediately obvious.  They may
  be associated with  an automorphism between the star
  product and the commutative product.\cite{Jurco:2000fb}}  \beqa {\cal D}_\theta\phi&=&\partial\phi\biggl(1-\frac
12\theta^{-1}\partial\theta\bar\partial\theta\biggr) -\frac 12\; \theta
\partial^2\bar\partial\phi  +\cdot\cdot\cdot\cr & &\cr  \bar{\cal D}_\theta\phi&=&\bar\partial\phi\biggl(1-\frac
12\theta^{-1}\partial\theta\bar\partial\theta\biggr) -\frac 12\; \theta
\bar \partial^2\partial\phi  +\cdot\cdot\cdot \eeqa
The result is quite simple when compared to previous approaches\cite{Pinzul:2005nh}.
  From the lowest order term we recover the
scalar field action in the conformal gauge,
$S_\phi^0\;$.  Conformal invariance is broken  by the
noncommutative corrections.   The lowest order corrections can be
expressed in terms of  the determinant of the classical metric using
(\ref{clsclmsr}) and 
(\ref{xpnsnfthta}).  The field equation following from
variations of $\phi$ in (\ref{frdrcrtdactnsc})  gives the conservation law
\be \bar\partial {\cal I}_\phi + \partial \bar{\cal I}_\phi = 0\;,\label{nccrtstsft}\ee
with currents
\beqa {\cal I}_\phi&=& \partial\phi
(1-\theta^{-1}\partial\theta\bar\partial\theta)-\frac 12 [
\theta\partial^2\bar\partial\phi +
\partial\bar\partial(\theta\partial\phi)]+\cdot\cdot\cdot\cr &
&\cr
\bar{\cal I}_\phi&=& \bar\partial\phi
(1-\theta^{-1}\partial\theta\bar\partial\theta)-\frac 12 [
\theta\bar \partial^2\partial\phi +
\partial\bar\partial(\theta\bar\partial\phi)]+\cdot\cdot\cdot\label{ncccrnts}  \eeqa

For gauge theories we can use
 (\ref{afsvt}) to compute lowest order  corrections to
 the field strength ${\cal F}$ in (\ref{smblsfs}) 
\be {\cal F} = \frac\theta\kbar\biggl( {\cal F}_M -\frac i2 (\partial\theta\bar\partial^2{\cal A}
 -\bar\partial\theta\partial^2\bar{\cal
 A})+\cdot\cdot\cdot\biggr)\;,\label{crfstr}\ee where ${\cal F}_M $ is
 the Moyal-Weyl field strength (\ref{fsmlwl}).  As a result we have
 corrections to the noncommutative plane limit, as well as the to
 commutative limit.  For the latter, note that there are terms of  order $\kbar$
 in ${\cal F}_M$, as well as in $\theta(z,\bar z)$ and its    derivatives. 
The lowest order  noncommutative corrections to the flux
  can  be read off from (\ref{gnrlflx}).  Using (\ref{twntefv}) and
 (\ref{twntesxv}) one gets the boundary terms
\beqa  \Phi_\sigma &=&\int_{\partial \sigma} \biggl\{dz \biggl(
{\cal A}-i\kbar {\cal A} \partial \bar{\cal A} +\frac 12 \partial
 \theta \bar\partial {\cal A}+\cdot\cdot\cdot\biggr)\cr & &\cr& &\quad
+\;\;d\bar z  \biggl(
\bar{\cal A}+i\kbar  \bar{\cal A}\bar \partial {\cal A} +\frac 12\bar \partial
 \theta \partial\bar {\cal A}+\cdot\cdot\cdot\biggr)\biggr\}\;\label{ctdncflz}
\eeqa 
Corrections to the gauge field  action are
\beqa S_f&=&\frac i{4\kbar}\int  dz\wedge d\bar z\;\theta\;
 \;\biggl\{{\cal F}_M^2(1+\frac 12 \partial\bar\partial\theta)
 - i{\cal F}_M(\partial\theta\bar\partial^2{\cal A}
 -\bar\partial\theta\partial^2\bar{\cal A})\cr & &\cr& &\qquad\qquad\quad\qquad +\;\; \theta^{-1}\partial(\theta
 {\cal F}_M)\bar\partial(\theta {\cal F}_M)+\cdot\cdot\cdot\biggr\}\;\label{frdrcrtdactn}.
\eeqa When
 $\theta(z,\bar z)= \kbar$ we easily recover the 
 expression for the noncommutative plane. A Seiberg-Witten map should
 be utilized to  compare with the
 commutative case, which we do in sec 5.5.

\subsection{Gauge transformations}

Although the field strength transforms covariantly under gauge
transformations, this is not the case the potentials.  Here we
compute corrections to gauge transformations of   ${\cal A}$ and
$\bar{\cal A}$ from i) the commutative limit
 (\ref{gtcomth}) and ii) the noncommutative plane limit (\ref{gtmylwl}).  For this  write infinitesimal gauge variations as
\be \delta {\cal A} = \partial \Lambda +\Delta\qquad \delta \bar {\cal
  A} = \bar\partial \Lambda+\bar\Delta\;,\label{crsgtcomth}\ee
and substitute into (\ref{gtrofsgc}) using (\ref{crfstr}) to get
\beqa & &\bar\partial\;\biggl(\Delta +( \frac12 \partial\theta
 -i\kbar{\cal A})
 \partial\bar\partial\Lambda +\theta {\cal F}_M\partial\Lambda +
 i\kbar\bar{\cal A} \partial^2\Lambda \biggr) \cr & &\cr &-&
\partial\;\biggl(\bar\Delta +( \frac12\bar \partial\theta
 +i\kbar \bar{\cal A})
 \partial\bar\partial\Lambda +\theta  {\cal F}_M\bar\partial\Lambda -
 i\kbar {\cal A}\bar \partial^2\Lambda \biggr)\;=\; 0\;,
\eeqa  up to first order in $\kbar$.   Then  we can write down
$\Delta$ and $\bar\Delta$, up to the divergence of some arbitrary  function
$\sigma$ \beqa \Delta & =&-( \frac12 \partial\theta
 -i\kbar{\cal A})
 \partial\bar\partial\Lambda -\theta  {\cal F}_M\partial\Lambda -
 i\kbar\bar{\cal A} \partial^2\Lambda +\partial\sigma \cr & &\cr 
\bar\Delta &=&-( \frac12\bar \partial\theta
 +i\kbar \bar{\cal A})
 \partial\bar\partial\Lambda -\theta   {\cal F}_M\bar\partial\Lambda +
 i\kbar {\cal A}\bar \partial^2\Lambda +\bar\partial\sigma \;
\eeqa  The divergence of $\sigma$  can be absorb 
 in a re-definition of the gauge parameter  $\Lambda \rightarrow \hat
 \Lambda =\Lambda + \sigma$,  yielding
\beqa \delta {\cal A} &=&\partial \hat\Lambda -i[{\cal
 A},\hat\Lambda]_{\star_M}- \frac12 \partial\theta
 \partial\bar\partial\hat\Lambda +(\kbar -\theta ){\cal
   F}_M\partial\hat\Lambda +\cdot\cdot\cdot\label{ncgtA} \\ & &\cr
 \delta \bar {\cal
  A} &= &\bar\partial\hat \Lambda-i[\bar {\cal A},\hat\Lambda]_{\star_M}- \frac12\bar \partial\theta
 \partial\bar\partial\hat\Lambda +(\kbar -\theta ){\cal
 F}_M\bar\partial\hat\Lambda  +\cdot\cdot\cdot\label{ncgtAbr} \eeqa  
From this we easily recover the expressions for  limit i) by setting
$\theta(z,\bar z)= \kbar$=0, and limit ii) after setting
 $\theta(z,\bar z)= \kbar$ with $\hat \Lambda =\Lambda$ or $\sigma=0$.
 Gauge
  transformations close after including the first order corrections.  
If one goes beyond the leading order, the closure of gauge
transformations should put restrictions on $\sigma$.  
The corresponding gauge variations of ${\cal Z}$ and $\bar{\cal Z}$
are
\beqa \delta {\cal Z} &=& -i[{\cal
Z},\hat\Lambda]_{\star_M}- \frac12 \partial\theta
 \partial\bar\partial\hat\Lambda +(\kbar -\theta ){\cal
   F}_M\partial\hat\Lambda +\cdot\cdot\cdot\label{ncgtch} \\ & &\cr
 \delta \bar {\cal
  Z} &= &-i[\bar {\cal Z},\hat\Lambda]_{\star_M}- \frac12\bar \partial\theta
 \partial\bar\partial\hat\Lambda +(\kbar -\theta ){\cal
 F}_M\bar\partial\hat\Lambda  +\cdot\cdot\cdot\label{ncgtchbr} \eeqa  
  Setting ${\cal A}$ and $\bar{\cal A}$ equal to zero in
  (\ref{ncgtA}) and
  (\ref{ncgtAbr})  gives a first order expression for pure gauge potentials
\beqa {\cal Z}_{pg}& =&-\frac i \kbar \bar z +\partial \hat\Lambda - \frac12 \partial\theta
 \partial\bar\partial\hat\Lambda +\cdot\cdot\cdot \cr & &\cr \bar {\cal
  Z}_{pg}&=&\;\;\frac i \kbar \bar z+\bar\partial\hat \Lambda- \frac12\bar \partial\theta
 \partial\bar\partial\hat\Lambda +\cdot\cdot\cdot \;,\eeqa which
 generalizes the infinitesimal version of pure gauge solutions (\ref{prgg})
 to the noncommutative plane. 

\subsection{Seiberg-Witten map}

We  now construct the the lowest order  map from the  gauge theory written
on a coordinate patch $P_0$ of commutative  manifold  ${\cal M}_0$ to the
noncommutative gauge theory. (A more  geometrical treatment can be
found in \cite{Jurco:2000fb}.)  We write the noncommutative potentials
and gauge parameter as functions of the commutative ones, ${\cal
  A}[a,\bar a]$, $\bar {\cal
  A}[a,\bar a]$, $\hat\Lambda [\lambda, a, \bar a]$, and require that the right hand
sides of (\ref{ncgtA}) and (\ref{ncgtAbr}) be equal to 
$ {\cal A}[a +\partial \lambda,\bar a +\bar\partial \lambda]-{\cal
  A}[a,\bar a]$ and 
$\bar {\cal A}[a +\partial \lambda,\bar a +\bar\partial \lambda]-\bar {\cal
  A}[a,\bar a]$,
respectively.  The first order equation is then solved by
\beqa {\cal
  A}[a,\bar a]&=&a(1-\theta f) + \frac{ i\kbar} 2(
a\partial\bar a- \bar a\partial a) -\frac 12 \partial\theta\bar\partial
a \cr & &\cr\bar{\cal
  A}[a,\bar a]&=&\bar a(1-\theta f) - \frac{ i\kbar} 2(
\bar a\bar\partial a-  a\bar\partial\bar a) -\frac 12 \bar\partial\theta\partial\bar
a \cr & &\cr\hat\Lambda [\lambda, a, \bar a]&=&\lambda + \frac{i\kbar}2(
a\bar\partial \lambda  -\bar a \partial \lambda)\;,\label{gnrlswmp}\eeqa where $f$ is
the commutative curvature, $ i{f}= \bar\partial {a}-\partial\bar
{a}$. (\ref{gnrlswmp}) reduces to the standard Seiberg-Witten map to
the Moyal plane when $\theta=\kbar$.  The map to the
noncommutative curvature up to first order in $\kbar$ is
\be {\cal F}[a,\bar a]=\frac\theta\kbar\biggl( f (1 - \frac 12
\partial\bar\partial \theta ) +i \bar\partial(\theta f
a)-i\partial(\theta f\bar a)\biggr)\label{swofF}
\ee
Substituting into the noncommutative flux (\ref{ctdncflz})  gives, up to
first order in $\kbar$,  \beqa  \Phi_\sigma&=&\int_{\partial \sigma}
\biggl\{dz\; a(1-\theta f+\cdot\cdot\cdot)
+d\bar z \; \bar {a}(1-\theta f+\cdot\cdot\cdot)\biggr\}\;\cr& &\cr
&=& \Phi_\sigma^0 + \Phi_\sigma^1\; + \cdot\cdot\cdot\;,\label{crtdflx}
\eeqa 
where $\Phi_\sigma^0 $ is the commutative flux (\ref{stksthm}) and 
\be  \Phi_\sigma^1\;= -\;\kbar \int_{\partial \sigma}\theta_0 f
(dz\; a
\;+\;d\bar z \; \bar {a})\ee 
After substituting 
(\ref{swofF}) into the action (\ref{ncactnsmbl}), we get  the lowest
order correction to the commutative action (\ref{loactn})
\beqa S_f&=&\frac i{4\kbar}\int_\sigma  dz\wedge d\bar z \;\;\biggl(\theta{f}^2(1 - \frac 12
\partial\bar\partial \theta-\theta f )\; +\; \partial(\theta
 f)\bar\partial(\theta f)\biggr)\label{swactn}\\
& &\cr
&=&S_f^0+ S_f^1 +\cdot\cdot\cdot\;,\nonumber
\eeqa  where $S_f^0$ is the commutative Maxwell action  (\ref{loactn})
and \be S_f^1=\frac {i\kbar}{4}\int_\sigma  dz\wedge d\bar z \;\;\biggl({f}^2(\theta_1 - \frac 12\theta_0
\partial\bar\partial \theta_0-\theta_0^2 f )\; +\; \partial(\theta_0
 f)\bar\partial(\theta_0 f)\biggr)\;, \label{focrstcma}\ee and we have dropped boundary terms.  As in the commutative theory, there are
no propagating degrees of freedom.  The field equations
resulting from variations of $a$ and $\bar a$ in (\ref{swactn}) imply that \be
f (1 - \frac 12
\partial\bar\partial \theta-\frac 32\theta f ) - \partial\bar\partial(
\theta f) =\frac C\theta\;,\label{sweom} \ee $C$ being a constant.  If
we apply the expansion (\ref{xpnsnfthta}),
then (\ref{sweom}) gives the solution for the corrected field strength
\be f=f_0+f_1+\cdot\cdot\cdot\;,\label{slnfo}
\ee
where   the lowest
order agrees with the commutative solution $f_0=  {C_0}/{\theta_0}$ and 
\be f_1= \frac {\kbar C_0}{\theta_0}\biggl( \frac 12\partial\bar\partial
  \theta_0-
  \frac{\theta_1}{\theta_0}\biggr)\;,\ee
and we also  expanded the constant
\be C=
  C_0\kbar -\frac32 C_0^2\kbar^2+ \cdot\cdot\cdot \;\ee 
 From  (\ref{fpua}),  $ C_0$
was equal to the flux per unit area of the solution in the
commutative theory.  The flux per unit area remains a constant, but
its value gets shifted from the commutative result 
\be \frac{ \Phi_\sigma}{2\pi\kbar\;\int_\sigma d\mu(z,\bar z)} =
 C_0(1-\kbar C_0+\cdot\cdot\cdot)\;,\label{ncfpua}\ee
where we used the corrected flux (\ref{crtdflx}) and measure (\ref{crtdmsr}).
The action per unit area of the solution, $S_f= S_{f_0+f_1}^0 +S_{f_0}^1$, also remains a constant, its
 value being shifted  from the commutative result by the same  factor appearing in (\ref{ncfpua}),  
\be \frac{C_0^2}4 
 (1-\kbar C_0+\cdot\cdot\cdot)\;\ee
Since these shifts are small for small $\kbar$ we say that the solutions are stable
under the inclusion of noncommutative corrections.

\section{Noncommutative AdS${}^2$}
 
\setcounter{equation}{0} 

We now apply the results of the previous section to obtain the lowest order  
noncommutative corrections to the scalar and gauge  theory actions on
the Lobachevsky plane and show that the solution to the commutative
Maxwell action receives no such corrections.   For other approaches to the noncommutative
AdS${}^2$ and the
Lobachevsky plane,  see \cite{Fakhri:2003cu}, \cite{Madore:1999fi}.

\subsection{Lobachevsky plane}

 We first review the commutative theory.
Here we project the  AdS${}^2$ measure 
 to the disc, which then defines the Lobachevsky plane. 
We start with AdS${}^2$ embedded in  three-dimensional Euclidean space with
coordinates $x_i,\;i=0,1,2,$ and the constraint
\be x_0^2-x_1^2-x_2^2=1 \;,\qquad x_0\ge 1 \label{adstwo}  \ee  A natural Poisson
structure on it is
\be \{x_0,x_1\} = x_2\qquad \{x_2,x_0\} = x_1 \qquad \{x_1,x_2\} =-
x_0\;,\label{so21pb}\ee
as it  preserves the  $SO(2,1)$  symmetry.
  We parameterize the
Lobachevsky plane by complex coordinates $z$ and $\bar z$, with $0\le
|z|^2< 1$.   The projection  from  AdS${}^2$ to the disc is given by 
\be z =\frac{x_1 -ix_2}{x_0 +1}\qquad   \bar z =\frac{x_1 +ix_2}{x_0
  +1}\label{prjctntlp}\ee  
The Poisson brackets  (\ref{so21pb}) are projected to 
\be \{z,\bar z\} = -i\theta_0(|z|^2)\;,\qquad \theta_0(|z|^2)=\frac 12
(1-|z|^2)^2 \;,\label{thzfrlp} \ee with the associated measure  given by
(\ref{clsclmsr}).   The boundary $|z|=1$  corresponds to
$x_0$ and $\sqrt{x_1^2+ x^2_2}$ going to infinity in the  AdS${}^2$ space.   From
 (\ref{thzfrlp}) it follows that the
non-commutativity will tend to zero as  the boundary is approached   in the
 noncommutative version of the theory.  This is fortunate
because of the known difficulties associated with defining  boundaries in
noncommutative field
 theory.\cite{Pinzul:2001qh},\cite{Balachandran:2003vm} We note that
 the Lobachevsky plane differs
from the disc, and hence the noncommutative version differs from the
fuzzy disc\cite{Lizzi:2003ru}, since the metric and  $ \{z,\bar z\}$ are not constants in
the interior.

 The commutative
Maxwell action (\ref{loactn}) is
\be S^0_f=\frac i{8}\int  dz\wedge d\bar z \;\;
(1-|z|^2)^2 f^2\;, \label{cactnfrst} \ee and the  solutions of the corresponding
free field equations are of the form
\be f=\frac {2C_0}{(1-|z|^2)^2}\;, \label{cstslnlp}\ee
which   diverges as the boundary
$|z|^2=1$ is approached.   A possible gauge choice for the
potential one-form is
\be {\tt a} =iC_0\; \frac {{\bar z}dz - z d{\bar z}}
{1-|z|^2}\;  \ee  The flux going through a disc of radius
$r=|z|<1$ is $\Phi_r= 4\pi C_0/(1-\frac 1{r^2})$.

\subsection{Noncommutative  case}

For  noncommutative AdS${}^2$ we replace the embedding coordinates
$x_i, \;i=0,1,2,$ by hermitean operators ${\bf x}_i, \;i=0,1,2$, satisfying \be
{\bf x}_0^2-{\bf x}_1^2-{\bf x}_2^2=1 \;, \label{ncadsthr}  \ee  with commutation relations 
\be [{\bf x}_0,{\bf x}_1] =i\kbar {\bf x}_2\qquad [{\bf x}_2,{\bf
  x}_0] =i\kbar {\bf x}_1 \qquad [{\bf x}_1,{\bf x}_2] =-i\kbar
{\bf x}_0\;\label{so21cr}\ee  Thus the $SO(2,1)$ symmetry of the
commutative theory is undeformed. Here, of course, there are no
finite dimensional unitary representations of the algebra.

 Next we write down an operator analogue
of the projection (\ref{prjctntlp}) to the Lobachevsky plane. Up to
operator ordering ambiguities we have
\be {\bf z} =({\bf x}_1-i {\bf x}_2)({\bf x}_0+1 )^{-1} \;,\qquad
   {\bf z}^\dagger =({\bf x}_0+1 )^{-1} ({\bf x}_1+i {\bf x}_2)\;,
   \label{prjctntnclp} \ee which is nonsingular if $-1$ is not in the
   spectrum of ${\bf x}_0$.   We can find the commutation relations
   for ${\bf z}$ and $ {\bf z}^\dagger$ in a manner similar  to what
   was done for the projective coordinates of the  fuzzy sphere\cite{Alexanian:2000uz}.
   For this define
  $\chi^{-1} ={1\over 2} (1+{\bf x}_0)$, which from (\ref{so21cr})
   commutes with $|{\bf z}|^2={\bf z}  {\bf z}^\dagger $.  After  some
   algebra one can show that 
\be \Theta( {\bf
  z},{\bf z}^\dagger) = \kbar \chi \biggl(1-|{\bf z}|^2 -{\chi\over 2}  (1+{\kbar\over 2}|
{\bf z}|^2 ) \biggr)  \;,\label{zzdag}  \ee   and from the constraint
   (\ref{ncadsthr}) one also has 
\be {\bf z}\chi^{-2} {\bf z }^\dagger + \chi^{-1}  {\bf z }^\dagger {\bf z}
 \chi^{-1} -2 \chi^{-1}(\chi^{-1} - 1) = 0 \label{chizsqr} \ee
Using the commutation relations it is then possible to solve for
   $\chi$ as a function of $|{\bf z}|^2$, and hence write $\Theta( {\bf
  z},{\bf z}^\dagger)$  as a function of just
   $|{\bf z}|^2$, and of course, the commutativity parameter $\kbar$.
   We won't write down the exact expression, but instead give the
   expansion up to first order  in  $\kbar$.
Actually, we only need the zeroth order expression for  $\Theta( {\bf
  z},{\bf z}^\dagger)$, its symbol being the classical
   answer (\ref{thzfrlp})  times $\kbar$, to write down the first order corrections
   to  the real scalar field action (\ref{frdrcrtdactnsc})
 and the conserved currents  ${\cal I}_\phi $ and $\bar
   {\cal I}_\phi$ in  (\ref{ncccrnts}).
Substituting (\ref{thzfrlp}) in the latter gives 
$$ {\cal I}_\phi={\cal I}_\phi^0 +{\cal I}_\phi^1 +\cdot\cdot\cdot
   \;,\qquad $$  \beqa {\cal I}_\phi^0 
   &=&\partial \phi \;, \cr & &\cr
   {\cal I}_\phi^1 
   &=&-\frac\kappa 2\biggl\{ (6|z|^2 -1){\cal I}_\phi^0  -(1-|z|^2)(z\partial {\cal I}_\phi^0 + \bar z \bar\partial{\cal I}_\phi^0) +
(1-|z|^2)^2\partial\bar \partial{\cal I}_\phi^0\biggr\},
 \eeqa
 and $\bar {\cal I}_\phi$ is the
   complex conjugate.    It follows that the solutions to the commutative
   theory are not preserved in the lowest order  noncommutative theory.  If $\phi_0$ is a solution to the commutative
   theory, i.e. $ \partial \bar\partial \phi_0=0,$ then one can set \be\phi
   =\phi_0+ \phi_1\cdot\cdot\cdot \;,\ee
and, in principle, iteratively solve 
   for the noncommutative corrections.   Substituting into
   (\ref{nccrtstsft}) gives the following for the first order
   correction $\phi_1$:
  \be \partial\bar\partial\phi_1 
   =\frac\kappa 2\biggl\{3(z \partial\phi_0+ \bar z \bar
   \partial\phi_0)+ \frac 12 (z^2 \partial^2\phi_0+ \bar z^2 \bar
   \partial^2\phi_0) \biggr\}
 \ee

To obtain the leading noncommutative corrections to gauge theory we
need to compute  $\theta$ beyond zeroth order.  This is since
$\theta_1$ appears in the lowest order corrections to the Maxwell
action (\ref{focrstcma}).  We first expand $\chi$ and substitute in
(\ref{chizsqr}) to get
\be \chi = (1-|{\bf z}|^2)\biggl(1+ \frac \kbar 4 (1-3|{\bf z}|^2)+
   {\cal O}(\kbar^2)\biggr)\; \ee
Then from (\ref{zzdag})
\be\Theta({\bf z},{\bf
  z}^\dagger)= \frac \kbar{2} (1-|{\bf z}|^2)^2 \biggl(
   1-\frac \kbar 2|{\bf z}|^2
+
   {\cal O}(\kbar^2)\biggr)\;,\ee 
whose symbol  $\theta(z,\bar z)$
is   \be\theta(z,\bar z)=  \frac \kbar {2}
(1-z\star\bar z)_{\star }^2 \star  \biggl(
   1-\frac \kbar 2 z\star\bar z +   {\cal O}(\kbar^2)\biggr)\;,\ee
   where $\alpha^{ 2}_\star=\alpha\star\alpha$.  If we apply the
   expansion of the star product, given in (\ref{afsvt}), up to order  $\kbar$
 this leads to
\be\theta(z,\bar z)= \frac\kbar 2
(1-|z|^2)^2 - \frac{\kbar^2} 2
(1-|z|^2)^3    +{\cal
  O}(\kbar^3) \label{focttht}
  \;,\ee  and so $\theta_1$ in (\ref{xpnsnfthta}) is \be\theta_1(|z|^2)=-  
\frac 12 (1-|z|^2)^3\;\label{pseftht}\ee   This leads to  \be \frac1 2\partial\bar\partial
  \theta_0-
  \frac{\theta_1}{\theta_0}=\frac 12 \;,\label{svntwntn} \ee and as a result the commutative
   measure only  gets
   corrected at first order  by an overall constant factor
\be d\mu(z,\bar z)=\frac i {\pi\kbar}\;\frac{ dz\wedge d\bar z}{(1-|z|^2)^2}\; \biggl(1
 +\frac \kbar 2 + {\cal O}(\kbar^2)\biggr)
 \ee   If one assumes a noncommutative analogue of the relationship
 (\ref{clsclmsr}) between the measure and the metric,
   then the   components of the metric tensor also scale  by the factor $1
 +\frac \kbar 2$, up to first order.
  From (\ref{svntwntn}) it also  follows that the solution (\ref{cstslnlp}) to the  commutative
   gauge  theory is unchanged at this order; i.e.  (\ref{cstslnlp})  satisfies the first order corrected
   field equation (\ref{sweom}), although the constant
   $C_0$ is modified from its commutative value.  In terms of this
   shifted constant, the flux per unit area in (\ref{ncfpua}) then
   goes to
$  C_0 [1-\kbar (C_0+\frac 12)]$, and the action per unit area is
shifted by  the same small factor.

\section{Concluding Remarks}

\setcounter{equation}{0}

In section 6 we found that the flux per
 unit area of solutions to  noncommutative Maxwell theory gets shifted  from the
 commutative answer.
If  the result can be applied to the example of  monopoles
on the fuzzy sphere that would imply  a  shift in the
 magnetic monopole charge.   However, it is not straightforward to apply
 the techniques of section 6 to the case of the fuzzy sphere.   Among
 the problems are: a) As mentioned earlier,  nonlinear
 coherent states $|z>$ satisfying
 (\ref{zee}) are not readily available for fuzzy manifolds.
 Alternatives states  have been found for the fuzzy sphere, and they lead to a
 modified star product and integration measure.\cite{Pinzul:2005nh},\cite{Alexanian:2000uz} 
b)  The Seiberg-Witten map is problematic  in the case of fuzzy
 gauge theories
 since  mapping from the noncommutative theory to the commutative one  would mean connecting  finite dimensional spaces with an
 infinite dimensional one.\cite{Grimstrup:2003rd}
c) A constraint must be introduced to insure that off block-diagonal
 matrix components of the potential remain non-dynamical when $\kbar$
 is small.  This
 constraint was necessary in the full noncommutative theory in order
 to obtain solutions with nonvanishing magnetic charge.  We also
 found that the    off block-diagonal
 matrix components should not vanish if we were to recover a finite
 value for the magnetic charge $g_0$ in the limit. 
If a proper treatment does  reveal a shift in the magnetic charge due
 to noncommutative corrections, one
 expects that
 there should also be a shift in the Dirac quantization condition.  For this it  would be useful to  analyze the
quantum mechanics of a particle in
a general monopole background  (\ref{qncmnplsns}) (similar to what was
 carried out in \cite{Karabali:2001te}).  This would require finding
 a gauge invariant coupling to the noncommutative potentials.  Constraints
should then be found  on the charges which reduce to the Dirac quantization condition in the
commutative limit.   

In the approach taken here we took advantage of the K\"ahler structure
in two dimensions to construct the fundamental
commutator (\ref{nclgbr}) from the determinant of the metric
tensor on $P_0$. As the commutative gauge theory action depends  on the metric
only through  its
determinant, no other ingredient was needed to write down the
noncommutative version of the theory.   This was not the case for the scalar field
theory.  There we chose a space-time gauge to get rid of additional
degrees of freedom in the metric.  More constraints will have to be imposed
in order to generalize this approach to  other field theories.  In
particular, frame fields and spin connections must be dealt with in
the case of fermions, and also for gravity.  Concerning the latter,
the non-commutativity was taken to be non-dynamical, and in fact a
constant, in most previous treatments of noncommutative
gravity.\cite{Aschieri:2005yw},\cite{Banados:2001xw},\cite{Cacciatori:2002gq},\cite{Pinzul:2005ta}  However, here since the
non-commutativity is connected to the metric tensor, a  consistent approach would mean  that $\Theta( {\bf
  z},{\bf z}^\dagger)$ should be elevated to a dynamical field, which should
lead to a more realistic model for quantum gravity (at least in two dimensions).

\bigskip
\bigskip

{\parindent 0cm{\bf Acknowledgment}}                                         
 
The author is very grateful to A. Pinzul for
discussions.


\bigskip
\bigskip
 

\begin{thebibliography}{99}

\bibitem{Balachandran:2005ew}
  A.~P.~Balachandran, S.~Kurkcuoglu and S.~Vaidya,
  %``Lectures on fuzzy and fuzzy SUSY physics,''
  arXiv:hep-th/0511114.

\bibitem{Fosco:2004yz}
  C.~D.~Fosco and G.~Torroba,
  %``Noncommutative theories and general coordinate transformations,''
  Phys.\ Rev.\ D {\bf 71}, 065012 (2005).

\bibitem{Gayral:2005ih}
  V.~Gayral, J.~M.~Gracia-Bondia and F.~Ruiz Ruiz,
  %``Position-dependent noncommutative products: Classical construction and
  %field theory,''
  Nucl.\ Phys.\ B {\bf 727}, 513 (2005).


\bibitem{Balachandran:1983pc}
  A.~P.~Balachandran, G.~Marmo, B.~S.~Skagerstam and A.~Stern,
  %``Gauge Symmetries And Fiber Bundles: Applications To Particle Dynamics,''
  Lect.\ Notes Phys.\  {\bf 188}, 1 (1983);
  A.~Berard and H.~Mohrbach,
  %``Monopole in momentum space in noncommutative quantum mechanics,''
  Phys.\ Rev.\ D {\bf 69}, 127701 (2004);
 P.~A.~Horvathy, L.~Martina and P.~C.~Stichel,
  %``Enlarged Galilean symmetry of anyons and the Hall effect,''
  Phys.\ Lett.\ B {\bf 615}, 87 (2005).

\bibitem{Pinzul:2005nh}
  A.~Pinzul and A.~Stern,
  %``A perturbative approach to fuzzifying field theories,''
  Nucl.\ Phys.\ B {\bf 718}, 371 (2005).


\bibitem{Behr:2003qc}
  W.~Behr and A.~Sykora,
  %``Construction of gauge theories on curved noncommutative spacetime,''
  Nucl.\ Phys.\ B {\bf 698}, 473 (2004).


\bibitem{Grosse:1992bm}
  H.~Grosse and J.~Madore,
  %``A Noncommutative version of the Schwinger model,''
  Phys.\ Lett.\ B {\bf 283}, 218 (1992).


\bibitem{Klimcik:1997mg}
  C.~Klimcik,
  %``Gauge theories on the noncommutative sphere,''
  Commun.\ Math.\ Phys.\  {\bf 199}, 257 (1998).

\bibitem{Grosse:1998da}
  H.~Grosse and P.~Presnajder,
  %``A treatment of the Schwinger model within noncommutative geometry,''
  Lett.\ Math.\ Phys.\ {\bf  46}, 61 (1998).

\bibitem{Presnajder:1999ky}
  P.~Presnajder,
  %``The origin of chiral anomaly and the noncommutative geometry,''
  J.\ Math.\ Phys.\  {\bf 41}, 2789 (2000); 
  %``Gauge fields on the fuzzy sphere,''
  Mod.\ Phys.\ Lett.\ A {\bf 18}, 2431 (2003).




\bibitem{Iso:2001mg}
  S.~Iso, Y.~Kimura, K.~Tanaka and K.~Wakatsuki,
  %``Noncommutative gauge theory on fuzzy sphere from matrix model,''
  Nucl.\ Phys.\ B {\bf 604}, 121 (2001).

\bibitem{Karabali:2001te}
  D.~Karabali, V.~P.~Nair and A.~P.~Polychronakos,
  %``Spectrum of Schroedinger field in a noncommutative magnetic monopole,''
  Nucl.\ Phys.\ B {\bf 627}, 565 (2002).


\bibitem{Balachandran:2003ay}
  A.~P.~Balachandran and G.~Immirzi,
  %``The fuzzy Ginsparg-Wilson algebra: A solution of the fermion doubling
  %problem,''
  Phys.\ Rev.\ D {\bf 68}, 065023 (2003).

\bibitem{Steinacker:2003sd}
  H.~Steinacker,
  %``Quantized gauge theory on the fuzzy sphere as random matrix model,''
  Nucl.\ Phys.\ B {\bf 679}, 66 (2004).



\bibitem{Baez:1998he}
  S.~Baez, A.~P.~Balachandran, B.~Ydri and S.~Vaidya,
  %``Monopoles and solitons in fuzzy physics,''
 Commun.\ Math.\ Phys.\  {\bf 208}, 787 (2000).




\bibitem{Alexanian:2000uz}
  G.~Alexanian, A.~Pinzul and A.~Stern,
  %``Generalized Coherent State Approach to Star Products and Applications to
  %the Fuzzy Sphere,''
  Nucl.\ Phys.\ B {\bf 600}, 531 (2001).



\bibitem{Kontsevich:1997vb}
  M.~Kontsevich,
  %``Deformation quantization of Poisson manifolds, I,''
  Lett.\ Math.\ Phys.\  {\bf 66} (2003) 157.

\bibitem{mmsz}  V.I. Man'ko,  G. Marmo, E.C.G. Sudarshan, F. Zaccaria, 
Physica Scripta {\bf 55}, 528 (1997).

\bibitem{Seiberg:1999vs}
N.~Seiberg and E.~Witten,
%``String theory and noncommutative geometry,''
JHEP {\bf 9909}, 032 (1999).



\bibitem{Jayewardena:1988td}
  C.~Jayewardena,
  %``Schwinger Model On S(2),''
  Helv.\ Phys.\ Acta {\bf 61}, 636 (1988).


\bibitem{Morariu:2005pv}
  B.~Morariu and A.~P.~Polychronakos,
  %``Fractional quantum Hall effect on the two-sphere: A matrix model
  %proposal,''
  Phys.\ Rev.\ D {\bf 72}, 125002 (2005).





  


\bibitem{Asakawa:1999cu}
T.~Asakawa and I.~Kishimoto,
%``Comments on gauge equivalence in noncommutative geometry,''
JHEP {\bf 9911}, 024 (1999);
B.~Jurco, L.~Moller, S.~Schraml, P.~Schupp and J.~Wess,
%``Construction of non-Abelian gauge theories on noncommutative spaces,''
Eur.\ Phys.\ J.\ C {\bf 21}, 383 (2001);
D.~Brace, B.~L.~Cerchiai, A.~F.~Pasqua, U.~Varadarajan and B.~Zumino,
%``A cohomological approach to the non-Abelian Seiberg-Witten map,''
JHEP {\bf 0106}, 047 (2001); D.~Brace, B.~L.~Cerchiai and B.~Zumino,
%``Non-Abelian gauge theories on noncommutative spaces,''
Int.\ J.\ Mod.\ Phys.\ A {\bf 17S1}, 205 (2002);
G.~Barnich, F.~Brandt and M.~Grigoriev,
%``Seiberg-Witten maps and noncommutative Yang-Mills theories for  arbitrary
%gauge groups,''
JHEP {\bf 0208}, 023 (2002).

\bibitem{Grimstrup:2003rd}
J.~M.~Grimstrup, T.~Jonsson and L.~Thorlacius,
%``The Seiberg-Witten map on the fuzzy sphere,''
JHEP {\bf 0312}, 001 (2003);  J.~M.~Grimstrup,
%``Noncommutative Coordinate Transformations And The Seiberg-Witten Map,''
Acta Phys.\ Polon.\ B {\bf 34}, 4855 (2003).

\bibitem{Pinzul:2004tq}
  A.~Pinzul and A.~Stern,
  %``Scale transformations on the noncommutative plane and the  Seiberg-Witten
  %map,''
  Int.\ J.\ Mod.\ Phys.\ A {\bf 20}, 5871 (2005).









  


\bibitem{Pinzul:2001my}
  A.~Pinzul and A.~Stern,
  %``A new class of two-dimensional noncommutative spaces,''
  JHEP {\bf 0203}, 039 (2002).


\bibitem{Jurco:2000fb}
  B.~Jurco and P.~Schupp,
  %``Noncommutative Yang-Mills from equivalence of star products,''
  Eur.\ Phys.\ J.\ C {\bf 14}, 367 (2000).

\bibitem{Fakhri:2003cu}
  H.~Fakhri and A.~Imaanpur,
  %``Dirac operator on noncommutative AdS(2),''
  JHEP {\bf 0303}, 003 (2003).

\bibitem{Madore:1999fi}
  J.~Madore and H.~Steinacker,
  %``Propagator on the h - deformed Lobachevsky plane,''
  J.\ Phys.\ A {\bf 33}, 327 (2000).

\bibitem{Pinzul:2001qh}
  A.~Pinzul and A.~Stern,
  %``Absence of the holographic principle in noncommutative Chern-Simons
  %theory,''
  JHEP {\bf 0111}, 023 (2001).

\bibitem{Balachandran:2003vm}
  A.~P.~Balachandran, K.~S.~Gupta and S.~Kurkcuoglu,
  %``Edge currents in non-commutative Chern-Simons theory from a new matrix
  %model,''
  JHEP {\bf 0309}, 007 (2003).

\bibitem{Lizzi:2003ru}
  F.~Lizzi, P.~Vitale and A.~Zampini,
  %``The fuzzy disc,''
  JHEP {\bf 0308}, 057 (2003);  Mod.\ Phys.\ Lett.\ A {\bf 18}, 2381
  (2003);  JHEP {\bf 0509}, 080 (2005).

\bibitem{Aschieri:2005yw}
  P.~Aschieri, C.~Blohmann, M.~Dimitrijevic, F.~Meyer, P.~Schupp and J.~Wess,
  %``A gravity theory on noncommutative spaces,''
  Class.\ Quant.\ Grav.\  {\bf 22}, 3511 (2005);  P.~Aschieri, M.~Dimitrijevic, F.~Meyer and J.~Wess,
  %``Noncommutative geometry and gravity,''
  arXiv:hep-th/0510059.

\bibitem{Banados:2001xw}
  M.~Banados, O.~Chandia, N.~E.~Grandi, F.~A.~Schaposnik and G.~A.~Silva,
  %``Three-dimensional noncommutative gravity,''
  Phys.\ Rev.\ D {\bf 64}, 084012 (2001).


\bibitem{Cacciatori:2002gq}
  S.~Cacciatori, D.~Klemm, L.~Martucci and D.~Zanon,
  %``Noncommutative Einstein-AdS gravity in three dimensions,''
  Phys.\ Lett.\ B {\bf 536}, 101 (2002).


\bibitem{Pinzul:2005ta}
  A.~Pinzul and A.~Stern,
  %``Noncommutative AdS(3) with quantized cosmological constant,''
  Class.\ Quant.\ Grav.\  {\bf 23}, 1009 (2006).

\end{thebibliography}
\end{document}